\documentclass[lettersize,journal]{IEEEtran}
\usepackage{amsmath,amsfonts}
\usepackage[english]{babel} 

\usepackage{algorithmic}
\usepackage{algorithm}

\usepackage{array}
\usepackage[caption=false,font=normalsize,labelfont=sf,textfont=sf]{subfig}
\usepackage{textcomp}
\usepackage{stfloats}
\usepackage{url}
\usepackage{verbatim}

\usepackage{cite}
\usepackage{diagbox}
\usepackage{enumerate}
\usepackage{threeparttable}
\usepackage{csquotes}
\usepackage{tabularx}

\usepackage{caption}

\usepackage{float} 
\usepackage{listings} 
\usepackage{arydshln} 
\usepackage{multirow}
\usepackage{multicol}
\usepackage{csquotes}
\usepackage{graphicx}
\usepackage{xcolor}

\begin{document}

\title{DTHA: A Digital Twin-Assisted Handover Authentication Scheme for 5G and Beyond}

\author{Guanjie Li, Tom H. Luan,  \emph{Senior Member, IEEE}, Chengzhe Lai, \emph{Member, IEEE},  \\Jinkai Zheng, \emph{Member, IEEE}, and Rongxing Lu, \emph{Fellow, IEEE}
\thanks{ 
This work was supported in part by the National Nature Science Foundation of China (NSFC), No. 62171352, and in part by the Program of China Scholarship Council (CSC), No. 202306960051. (\textit{Corresponding author: Tom H. Luan and Chengzhe Lai)}

Guanjie Li  is with the School of Cyber Engineering, Xidian University, Xi'an, 710071, China. Email: sinleced@msn.cn.

Tom H. Luan and Jinkai Zheng are with the School of Cyber Science and  Engineering, Xi'an Jiaotong University, Xi'an, 710049, China. Email:  tom.luan@xjtu.edu.cn; jkzheng@stu.xidian.edu.cn.

Chengzhe Lai is with the School of Cyberspace Security, Xi'an University of Posts and Telecommunications, Xi'an, 710061, China. Email: lcz\_xupt@163.com.

Rongxing Lu is with the School of Computing, Queen's University, Kingston, Ontario, Canada.  E-mail: rxlu@ieee.org.

}

}

\markboth{Journal of \LaTeX\ Class Files,~Vol.~14, No.~8, August~2021}%
{Shell \MakeLowercase{\textit{et al.}}: A Sample Article Using IEEEtran.cls for IEEE Journals}

\maketitle

\begin{abstract}
With the rapid development and extensive deployment of the fifth-generation wireless system (5G), it has achieved ubiquitous high-speed connectivity and improved overall communication performance. Additionally, as one of the promising technologies for integration beyond 5G, digital twin in cyberspace can interact with the core network, transmit essential information, and further enhance the wireless communication quality of the corresponding mobile device (MD). However, the utilization of millimeter-wave, terahertz band, and ultra-dense network technologies presents urgent challenges for MD in 5G and beyond, particularly in terms of frequent handover authentication with target base stations during faster mobility, which can cause connection interruption and incur malicious attacks. To address such challenges in 5G and beyond, in this paper, we propose a secure and efficient handover authentication scheme by utilizing digital twin. Acting as an intelligent intermediate, the authorized digital twin can handle computations and assist the corresponding MD in performing secure mutual authentication and key negotiation in advance before attaching the target base stations in both intra-domain and inter-domain scenarios. In addition, we provide the formal verification based on BAN logic, RoR model, and  ProVerif, and informal analysis to demonstrate that the proposed scheme can offer diverse security functionality. Performance evaluation shows that the proposed scheme outperforms most related schemes in terms of signaling, computation, and communication overheads.

\end{abstract} 


\section{Introduction}
\IEEEPARstart 
{O}{ver} the past few decades, the rapid evolution of mobile devices (MDs) have transformed society, enabling unprecedented access to digital services through smartphones, wearables, and connected vehicles. The exponential growth of these devices has generated massive data traffic, creating unprecedented demands for reliable wireless communications \cite{kim2024role}. To meet these requirements, the current fifth-generation wireless system (5G) leverages mmWave spectrum and other key technologies to achieve 50 Mbps data rates and offer other impressive features \cite{3gpp.36.331}\cite{huang2020secure}. Beyond 5G (B5G), the next-generation communication system is envisioned to use the ultra-wide terahertz (THz) band to further expand bandwidth and throughput for carrying the access of massive devices in the near future \cite{yang2022terahertz} \cite{wang2023road}. To overcome transmission loss introduced by mmWave and THz, additional, ultra-dense networks (UDN) have been proposed by densely deploying a large number of small cells to improve spectrum efficiency \cite{sharma2021resource}.

While 5G and B5G can enhance the quality of communication, they also present additional mobility management challenges in high-speed mobility scenarios, such as high-speed railways and autonomous driving \cite{lai2022novel} \cite{park2024mobility}. The dense deployment of short-range base stations forces mobile devices to perform frequent handover authentication every few seconds to maintain continuous wireless connectivity.  Additionally, the openness of the wireless channel makes the handover process susceptible to malicious attacks, such as eavesdropping and replay attacks \cite{ma2019ftgpha}, creating vulnerabilities that can lead to unauthorized access and compromised privacy. 
Although researchers have proposed various solutions leveraging advanced technologies like software-defined networks (SDN) and blockchain to enhance handover security and efficiency \cite{zhang2025lshsc} \cite{yazdinejad2019blockchain},  a key limitation of most existing schemes lies in adopting a reactive approach that only initiates handover authentication when mobile devices enter target base stations' coverage areas and reach the handover signal threshold. This device-initiated, network-controlled handover mechanism can lead to excessive latency and overhead, disrupting the stable and continuous connectivity of mobile devices to the network through base stations \cite{park2024mobility}  \cite{fang2024decentralized}.

As one of the promising technologies to be integrated into B5G, the emerging digital twin paradigm has garnered attention from both academia and industry \cite{khan2022digital}. A digital twin refers to the virtual agent of mobile device deployed in cyberspace \cite{li2023breaking} \cite{gautam2024blockchain}. It receives and synchronizes multi-dimensional status data transmitted from mobile device, and employs artificial intelligence algorithms for process and analysis to generate decision feedback, such as providing real-time path planning and navigation for autonomous vehicles \cite{hu2024d}. In B5G, digital twins can be used to further predict and optimize communication performance \cite{khan2022digital} \cite{3gppTR28915}. By monitoring and evaluating the wireless link status, such as signal strength, the digital twin can provide feedback to the core network when detecting performance degradation through the dedicated application programming interface (API). Based on this feedback, the core network takes timely optimization measures, such as wireless resource allocation, to improve the communication quality of mobile devices \cite{gao2023digital}.

This capability enables digital twin to assist mobile devices in proactive handover authentication, supporting the paradigm shift of mobility management in 5G and beyond \cite{zheng2024digital}. On the one hand, the digital twin can predict and select the upcoming target base stations connections by analyzing devices' movement trajectories, especially for autonomous vehicle. On the other hand, the digital twin can act as an intelligent intermediary to access the core network as well as the predicted target base stations, transmit essential handover authentication parameters, and burden complex computation on behalf of the corresponding mobile device in advance. Builiding on this foundation, in this paper, we propose a novel \underline{D}igital \underline{T}win-assisted \underline{H}andover \underline{A}uthentication scheme (DTHA) that achieves mutual authentication and key negotiation between mobile device and target base stations for 5G and beyond wireless network. By delegating authentication tasks to digital twin, the proposed scheme ensures handover security while significantly reducing authentication overhead  for mobile device. Specially, the main contributions are as follows:

\begin{itemize}

  \item \emph{Architecture}: We propose the first digital twin-assisted handover authentication architecture where the digital twin acts on behalf of the mobile device in operator-authorized cyberspace. With authorized access to the core network and base stations, this digital twin-driven approach transforms reactive authentication patterns by enabling proactive handover management and secure authentication parameter exchange,  reducing handover overhead in 5G and beyond networks.
  
  
  \item \emph{Design}: We design handover authentication protocols for both intra-domain and inter-domain scenarios.  The digital twin first obtains the delegation authority from the core network to act as a legitimate representative of the mobile device. Then, it proactively initiates handover authentication requests to target base stations and executes essential parameter transmission and complex computations on behalf of mobile device, enabling both the mobile device and target base stations to independently complete mutual authentication and key negotiation in advance.

  

  \item \emph{Validation}: We conduct comprehensive security validation through three formal methods, Burrows-Abadi-Needham (BAN) logic, Real-or-Random (RoR) model, and ProVerif, which proves our scheme achieves mutual authentication, key negotiation and session key security. The informal analysis demonstrates that the proposed scheme can resist various attacks. Furthermore,  comparative performance evaluations show that our scheme significantly outperforms most existing solutions in terms of signaling, computation, and communication overhead.
  
\end{itemize}

The organization of the rest of the paper is as follows. We first review the related works in section II. Then, we illustrate the system architecture, security threats and design goals in section III. We further detail the proposed scheme in section IV.  Security analysis and performance evaluation of the proposed scheme are presented in section V and section VI, respectively. Finally, we draw the conclusion of the paper in section VII.

\section{Related Work}
Handover authentication secures network access during user mobility by establishing mutual authentication between the mobile device and target access points, and negotiating session keys to protect against wireless vulnerabilities. The third generation partnership project (3GPP) has already standardized the 5G handover procedure in R16 \cite{3gpp.36.331}. However, it is vulnerable to suffering from different threats, such as de-synchronization attacks, replay attacks, etc \cite{huang2020secure}. Recent research has leveraged innovative technologies such as SDN, blockchain, and others to enhance the security and efficiency of handover authentication. We summarize some of these representative works as follows.

SDN separates the control plane and data plane in 5G, improving the flexibility and programmability of the network. Duan et al. \cite{duan2015authentication} showed that SDN can share user-specific secure context information (SCI) with predicted cells to enable seamless handover authentication.
 On this basis, Cao et al. \cite{cao2019cppha} proposed a pre-handover authentication mechanism where SDN forwards the device's session key and relevant information to the predicted next base station. 
 In \cite{lai2022novel}, SDN is employed to predict the moving path and inform the next base stations to initiate the handover procedure when the mobile device groups reach the signal threshold. Ma et al. \cite{ma2019ftgpha} proposed an SDN-based 5G group handover authentication scheme that generates and distributes session keys using 5G-AKA between base stations and mobile nodes. However,  the scheme \cite{ ma2019ftgpha} is limited to scenarios with fixed mobility trajectories, and the scheme \cite{lai2022novel} is vulnerable to replay attacks and man-in-the-middle attacks during the mobile device's handover request submission to the base station. In addition, these schemes \cite{lai2022novel, ma2019ftgpha} employ aggregate message authentication codes (AMAC), which introduces susceptibility to Denial of Service (DoS) attacks during the handover process.

 Blockchain is distributed ledgers that record transactions without tampering, which has been widely used in authentication management.  Zhang et al. \cite{zhang2019robust} and Liu et al. \cite{liu2024pecha} utilized blockchain to store chameleon values of mobile devices and base stations for quick handover.  However, Zhang et al.'s scheme \cite{zhang2019robust} is vulnerable to tracking attack and location privacy leakage, as the chameleon value can be computed by any receiver.  In response to this, Liu et al. \cite{liu2024pecha}  proposed periodically updating the chameleon values of mobile devices in the InterPlanetary File System (IPFS) to reduce the risk of being tracked. However, frequent handover of large-scale users could lead to excessive IPFS load, affecting data retrieval efficiency. Son et al. \cite{son2022design} designed a lightweight handover authentication scheme for vehicular networks where target RSUs retrieve vehicle information from the blockchain to enable efficient authentication and key agreement. Roy et al. \cite{roy2023blockchain} proposed utilizing RSUs as consortium blockchain nodes, storing vehicle handover authentication data as transactions on the blockchain, enabling RSUs to quickly verify vehicle identities and accomplish seamless handover authentication. Similarly, Sanjeev et al. \cite{10489855} proposed a dual-blockchain architecture to store vehicle and RSU registration information, leveraging smart contracts for cross-region handover authentication. However, in this scheme, blockchain consensus verification can introduce delays, potentially causing vehicle handover authentication failures in high-mobility scenarios. Fang et al. \cite{fang2024decentralized} used blockchain to protect vehicle authentication data in a zero-trust IoV environment, reducing handover authentication latency and utilizing smart contracts for reputation scoring to prevent malicious nodes from disrupting the process. However, attackers can create multiple fake ACs through Sybil attacks, enhancing the reputation of malicious nodes and allowing unauthorized vehicles to disguise themselves as legitimate to pass authentication.


 

In addition, several works apply digital signatures, physical layers, and other techniques to complete handover authentication. Gupta et al. \cite{gupta2018proxy} used proxy signature so that the base station and mobile device can mutually verify the proxy signature delegated by the core network to complete the authentication. However, it fails to effectively protect user privacy and is vulnerable to replay attacks, due to the MD information contained in the proxy certificate being publicly transmitted in communication channels. The scheme \cite{bi2025towards} introduces dynamic universal accumulators and sanitizable signatures to achieve privacy-preserving handover authentication. However, it fails to meet the requirements of Known Randomness Secrecy  and Key Escrow Freeness.  Yan et al. \cite{yan2022efficient} employed aggregate signatures to achieve handover authentication between vehicle groups and base stations and utilized a binary search to mitigate DoS attacks. However, the scheme fails to defend against an ephemeral secret leakage attack.  In \cite{cao2020lsaa}, Cao et al. employed extended Chebyshev chaotic maps to complete access authentication for mobile devices and massive machine-type communication devices in 5G, but this scheme requires key generation center to generate keys for MD and base stations, making it impossible to achieve key escrow freedom. \cite{roy2022fasthand} realized handover authentication with the cooperation of least $t$ out of $n$ base stations based $(t,n)$ threshold,  but it fails to consider node selection criteria for optimizing threshold signature security. In \cite{nyangaresi2022machine}, the authors designed a machine-learning protocol to select a target cell and guarantee both security and privacy during handovers in 5G, however, each handover authentication requires core network participation, which introduces additional communication and computation delays. Zhou et al. \cite{zhou2022handover} proposed using Automatic Dependent Surveillance-Broadcast (ADS-B) to collect side information for optimizing handover decisions in Aerial-Ground Vehicular Networks. However, it is vulnerable to jamming and spoofing attacks, affecting handover reliability.

\section{System model, Threat model, and Design Goals}
In this section, we first present the three-layered system model and describe the different entities that operate within each layer. We then discuss the threat model and identify the design goals in the proposed scheme.
\subsection{System Model}
\begin{figure}[t]
  \centering
  \includegraphics[width=8.4cm]{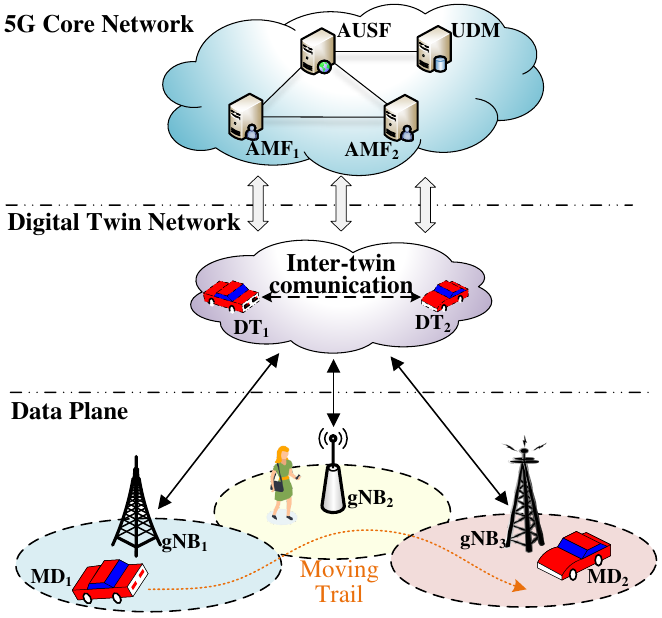}
  \caption{System Model}
  \label{fig:system}
\end{figure}

Fig. \ref{fig:system} illustrates the system model in the proposed scheme, which is divided into three layers based on their practical requirements and specific functions: the 5G core network (5GC), the digital twin network (DTN), and the data plane (DP), with each layer consisting of various entities denoted in bold. The details of each layer and their corresponding entities are described below.

\begin{itemize}
  \item  \textbf{5GC}: The 5G core network serves as the management center of the wireless network and plays a crucial role in network access, connection and routing. The network functions responsible for security and authentication in the 5GC primarily include the Access and Mobility Management Function (\textbf{AMF}), the Authentication Server Function (\textbf{AUSF}), and the Unified Data Management (\textbf{UDM}).  AUSF and  UDM are responsible for initial registration, identity management, and issuance of public-private key pairs in the wireless system. On the other hand,  AMF is responsible for mobility management and generating globe temporary identity (GUTI) for each mobile device (MD), which controls the handover of MDs between different base stations, ensuring that the 5GC provides secure and reliable wireless communication services.
  
  \item \textbf{DTN}: Digital twin network is the virtual cyberspace operated on dedicated cloud servers provided by telecom operators. Authorized MDs with service subscriptions in the 5GC can create and deploy corresponding digital twins (\textbf{DTs}) on the DTN. These DTs not only provide decision feedback for the  MDs via intra-twin communication but also communicate with each other to transmit useful data via inter-twin communication. Additionally, authorized DT on DTN can obtain the access methods of 5G network functions and base stations, enabling the establishment of connections and transmitting information through proprietary APIs, which can underpin valuable information for MD. In the proposed scheme, through trajectory analysis, the primary function of DTs is to pre-exchange essential parameters with the target base stations that the MD is expected to attach to along its route and further reduce the authentication overhead of MD during handover.
  
  
  \item \textbf{DP}: Data plane consists of mobile device (\textbf{MD}) and 5G base station \textbf{gNB}. In 5G, the communication between the gNB and the MD is bidirectional, with both entities transmitting and receiving data via a wireless channel.  On the one hand, MD such as smartphones, wearable devices and autonomous vehicles require access to the data network through gNB. The gNB maintains the connection with the AMF and is responsible for transmitting and forwarding data to AMF-authorized MD within its signal coverage area. Furthermore, to ensure wireless connection quality before leaving the current gNB, handover authentication with the new gNB in the mobile path is necessary due to the mobility of the MD.
\end{itemize}
\subsection{Threat Model}
In this paper, we adopt the Dolev-Yao (DY), Canetti-Krawczyk (CK) and extended CK (eCK) adversary models to identify potential attacks and vulnerabilities in our proposed handover authentication scheme involving MD, DT, gNB and AMF. Data is transmitted between the MD and DT/gNB over wireless channel, while the DT and gNB/AMF communicate over wired channel. Within these threat models, an adversary possesses multiple attack capabilities. Through the vulnerable wireless channel, the adversary can perform passive and active attacks, including eavesdropping, interception, replay, and message modification of transmitted data between entities. The adversary can also execute impersonation attacks by fraudulently assuming the identity of legitimate gNBs to deceive MDs into establishing connections. Furthermore, the adversary can acquire either the long-term secrets of the MD, DT, and gNB or their ephemeral secrets during the handover authentication phase. However, the simultaneous compromise of both long-term and ephemeral secrets is infeasible under the security model.


\subsection{Design Goals}
Our goal is to utilize the digital twin to provide a secure and efficient handover authentication for MD in 5G and beyond wireless networks. In order to address the aforementioned threats, the following design goals should be met:

\begin{itemize}
  \item Mutual Authentication: As the primary goal in the proposed scheme, mutual authentication between the MD and target gNB is essential to confirm the identity of both parties before granting handover access. In addition, acting as the representative of the MD, the DT should authenticate with the target gNBs and AMF to ensure its legitimacy and authorization.


  \item Key Negotiation: In the proposed scheme, to ensure the confidentiality and integrity of transmitted data over the public wireless channel, the MD and target gNB must negotiate a temporary session key during the handover authentication process.

  \item Anonymity: In the handover authentication phase, only the 5GC are permitted to disclose the true identity of the MD. Otherwise, the identity should be substituted with a pseudonym or concealed in ciphertext to prevent unauthorized disclosure by malicious attackers. 
  
  \item Unlinkability: It is impossible for an adversary to distinguish that two messages originate from the same MD when eavesdropping on the wireless channel in the handover phase.
  
  \item Traceability: When malicious behavior occurs during the handover authentication phase, such as an unauthorized DT attempting to gain access to the core network or the base station, it is essential to expose the true identity of the DT and the MD it represents.

  \item Perfect Forward/Backward Secrecy (PFS/PBS): PFS ensures that even if the private keys used by gNB and MD are compromised, an adversary cannot recover the historical temporary session keys. PBS means that the attacker cannot access future temporary session keys even if the current key is compromised.


\item Key Escrow Freedom (KEF): Each member's private key is kept by itself, there is no need for trusted third parties to distribute key materials in the authentication phase.

Ephemeral Secrets Leakage Resistance (ESL): Even if the ephemeral secret key is leaked to the adversary, the negotiated temporary session key between the target gNB and the MD remains confidential.


  \item Protocol Attack Resistance: The proposed scheme is designed to resist common protocol attacks, such as eavesdropping, forging attacks, impersonation attacks, man-in-the-middle attacks, and others.
\end{itemize}
\section{The proposed scheme}
In this section, we review the preliminaries employed in the scheme and then detail the digital twin-assisted handover authentication scheme in 5G.
\subsection{Preliminaries}
\subsubsection{Elliptic Curve Cryptography}
We adopt the elliptic curve cryptosystem (ECC) to ensure the security of our proposed scheme. Let $E/{F}_p$ denote an elliptic curve $E$ defined over a prime finite field ${F}_p$, where $p$ is a large prime number. Let $P$ be a generator point of $E$ with prime order $q$. The elliptic curve equation is defined as $E({F}_p): y^2 \equiv x^3 + ax + b\, mod\,{p}$, where $a,b \in {F}_p$ and $4a^3 + 27b^2 \not\equiv 0 \, mod\,{p}$.
\subsubsection{Mathematical Assumptions}

The security of the proposed scheme is based on the following mathematical assumptions:
\begin{itemize}
  \item Elliptic curve discrete logarithm problem (ECDLP): Given an additive group $G$ of prime order $q$ with generator point $P$ and $ xP \in G$, where $x \in {Z}_q^*$. It is computationally infeasible to determine the value of $x$ from $xP$.
  \item Elliptic curve computational Diffie-Hellman problem (ECDHP): Given an additive group $G$ of prime order $q$ with $P$ and two elements $xP, yP\in G$, where $x,y \in Z_q^*$. It is computationally infeasible to compute the $xyP$ with any polynomial time algorithm.
  \item Bilinear  Diffie–Hellman problem (BDHP): It is hard to distinguish between two elements $(g_1, g_2, g_T)$ and $(g_1, g_2, e(g_1, g_2))$ in two groups $G_1$ and $G_2$ that share a bilinear pairing $e: G_1 \times G_2 \rightarrow G_T$.
\end{itemize}
\subsection{Brief Description of the Scheme}
We begin by providing a brief description of our proposed scheme,  which consists of 5 phases: System Initialization, DT Creation, Access Delegation, Intra-AMF Handover and Inter-AMF Handover.

\textbf{System Initialization:} The 5GC is responsible for booting up and initializing the system, which includes issuing system parameters and generating partial public/private key pairs for AMF and gNB.

\textbf{DT Creation:} The creation of the DT on the DTN requires approval from the 5GC after the MD's request is submitted. Upon approval, the DT then has the ability to access the gNB and AMF to transmit the essential data and enable the added-value service for its MD after deployment. Additionally, to ensure secure communication, the DT and its MD should negotiate a long-term session key.

\textbf{Access Delegation:} In this phase, the DT applies for access delegation from the AMF using the token issued by the MD. Upon verifying the legality of the DT and the correctness of the token, the AMF issues the delegation to the DT, indicating its authorization to access all gNBs within the current domain.

\textbf{Intra-AMF Handover:} In this phase, DT predicts MD's movement trajectory and initiates handover requests to target gNBs via dedicated API. The process begins with mutual authentication between DT and target gNBs. DT then facilitates session key negotiation by transmitting necessary parameters between MD and target gNBs, reducing MD's authentication overhead. When MD enters the target gNB's coverage area, it can complete handover authentication and establish secure communication using the pre-computed session key.



\textbf{Inter-AMF Handover:} When the MD is about to move to a new gNB under another AMF control range, the DT notifies both the source AMF and MD to trigger inter-AMF handover authentication. The MD transmits the newly generated pseudonym identity to the DT through a secure wireless channel, while the source AMF sends essential security context information to the target AMF to generate new access delegation for the DT in advance. The subsequent mutual authentication and key agreement between the MD and target gNB follows the same process as intra-AMF handover.
\subsection{The Details of Proposed Scheme}
The main notations used in the proposed scheme and their corresponding descriptions are shown in Table I. The details of the scheme are as follows:
\begin{table}[ht]
\centering
\renewcommand{\arraystretch}{1.25}
\caption{Notations Used}
\begin{tabular}{l|l}
\hline
\textbf{Notation}           & \textbf{Description}                                                                                   \\ 
\hline

$MD_i$, $DT_j$ &        The mobile device and digital twin respectively\\
$SUPI_i$ & The subscription permanent identifier of mobile device\\
$GUTI_i$& The globally unique temporary identity of mobile device\\
$TID_i$ & The temporary identifier of mobile device in the gNB\\
$N_i$ & The nonce number\\
$TS_i$ & The current timestamp  \\
$q$  & The large prime number\\
$G$  & The additive cyclic group \\
$P$ & The generator of the additive cyclic group\\

$H(\cdot)$ & Collision-resistant cryptographic hash function\\
$KDF(\cdot)$ & Key derivation function\\
$pk$, $Y$                 & The public key pairs\\
$sk$, $x$                 & The private key pairs\\
 $k_{ij}$                  &  The long-term session key between MD and DT\\
 $k_{SEAF}$     & The security anchor key between MD and core network\\
 $k_{gNB}$      &  The session key between MD and gNB\\
 $\delta$      &  Authorized delegation \\
$||,\oplus$                & Concatenation and bitwise XOR operations respectively\\
\hline
\end{tabular}
\label{tab:math_symbols}
\end{table}

\subsubsection{\textbf{System Initialization}}
The  AUSF generates the system parameters and master key pair, while the AMF and gNB generate their respective public-private key pairs and obtain certified partial private keys from AUSF.
\begin{enumerate}[Step 1:]
\item AUSF: System Parameters\\
1a. AUSF initiates system initialization by taking a security parameter $\kappa$ as input. Subsequently, it generates a large prime number $p$ of $\kappa$ bits in length to establish the finite field $F_p$. \\
1b.  AUSF selects a generator $P$ of prime order $q$ on the  elliptic curve  $E({F}_p)$. It then defines an additive cyclic group $G$ generated by $P$ and a multiplicative cyclic group $G_T$ of order $q$ in the finite field.  AUSF defines a bilinear pairing map as $e: G \times G \rightarrow G_T$.\\
1c. AUSF selects cryptographic hash functions as $H_0:G\times \{0,1\}^*\rightarrow Z_q^*$, $H_1:G\rightarrow\{0,1\}^l$, $H_2:G\times \{0,1\}^*\rightarrow \{0,1\}^l$, $H_3:G\times G_T\times \{0,1\}^*\rightarrow \{0,1\}^l$, $H_4:\{0,1\}^*\rightarrow\{0,1\}^l$, and $H_5:\{0,1\}^*\rightarrow Z_q^*$.\\
1d. AUSF randomly chooses $s\in Z_q^*$ as the master private key and computes the corresponding public key $pk_{pub}=s\cdot P$. Finally, AUSF publishes the system public parameters $\{q,e,P,G,G_T,H_0,H_1,H_2,H_3,H_4,H_5,pk_{pub}\}$ while keeping $s$ secret.
\item AMF\slash gNB $\rightarrow$ AUSF: $(pk_a,ID_a)$/$(pk_g,ID_g)$\\
2a. AMF and gNB randomly select their private keys $sk_a, sk_g \in {Z}_q^*$ and compute the corresponding public keys $pk_a=sk_a\cdot P$ and $pk_g=sk_g \cdot P$ separately.\\
2b.  AMF and gNB transmit $(ID_a, pk_a)$ and $(ID_g, pk_g)$ to AUSF  through secure communication channel, where $ID_a$ and $ID_g$ represent the unique identifiers of AMF and gNB respectively.
\item AUSF $\rightarrow$ AMF\slash gNB: $(x_a,Y_a)$/$(x_g,Y_g)$\\
3a. Upon receiving messages from AMF and gNB, AUSF randomly selects $y_a,y_g\in Z_q^*$ and computes $Y_a=y_a\cdot P$ and $Y_g=y_g\cdot P$.\\
3b. AUSF uses the master private key $s$ to compute partial private keys $x_a=y_a+s\cdot H_0(ID_a,pk_a)\,mod\,q$ and $x_g=y_g+s\cdot H_0(ID_g,pk_g)\,mod\,q$.\\
3c. AUSF securely transmits $(x_a,Y_a)$ and $(x_g,Y_g)$ to AMF and gNB respectively.
\item AMF\slash gNB: $(pk_a,Y_a)$/$(pk_g,Y_g)$\\
4a. After receiving messages from AUSF, AMF verifies the correctness by checking the equation $x_{a}\cdot P \stackrel{?}{=} Y_{a} + H_0(ID_{a}, pk_{a})\cdot pk_{pub}$. Similarly, gNB performs the corresponding verification process.\\
4b. Upon successful verification, AMF and gNB securely store their complete private key pairs $(sk_{a}, x_{a})$ and $(sk_{g}, x_{g})$ respectively and publish  $(ID_a, pk_{a}, Y_{a})$ and $(ID_g, pk_{g}, Y_{g})$ to the network.
\end{enumerate}
\subsubsection{\textbf{DT Creation}}
The MD generates its key pair and requests a DT creation, after which the DT is initialized with corresponding keys and establishes a secure channel with the MD. The details are as follows.

\begin{enumerate}[Step 1:]
    \item $MD_i\rightarrow 5GC$: Subscription Request $(SUPI_i,pk_i)$\\
    1a. The $MD_i$ first randomly selects $sk_i\in Z_q^*$ as its private key, and calculates $pk_i=sk_i\cdot P$ as its public key.\\
    1b. The $MD_i$ sends subscription request $req_i = (SUPI_i, pk_i)$ to the AUSF and UDM in the 5GC, where $SUPI_i$ is the subscription permanent identifier, to apply for creating a corresponding virtual twin $DT_j$ in the DTN.
    \item $DT_j\leftrightarrow  MD_i$: $k_{ij}$\\
    2a. After UDM verifies the request information from $MD_i$ and confirms it meets the qualification conditions (such as completed payment), the NEF network function coordinates to allocate computing resources on dedicated cloud servers, creating and deploying its digital twin $DT_j$.\\
    2b. When $DT_j$ is successfully activated, $DT_j$ randomly selects $sk_{j}\in Z_q^*$ as its private key, and calculates the corresponding public key $pk_{j}=sk_{j}\cdot P$.\\
    2c. AUSF randomly selects $u_j\in Z_q^*$ and calculates $U_j=u_j\cdot P$. Additionally, AUSF uses the system master key to generate an anonymous identity for $DT_j$ as $ID_{j}=SUPI_i\oplus H_1(sU_j)$. UDM sends $(ID_j, pk_j)$ to AMF and gNB in the 5G network, and sends $(SUPI_i,pk_i)$ to AMF. Furthermore, UDM stores $(ID_j, U_j)$ in the local security database for identity tracing in subsequent processes.\\
    2d. After $DT_j$ is successfully deployed to the DTN, it accesses 5GC and gNB through dedicated API interfaces \cite{khan2022digital}. Subsequently, $DT_j$ negotiates a long-term session key $k_{ij}$ with $MD_i$, to achieve secure data transmission via intra-twin channel\cite{li2021seccdv}.
\end{enumerate}

\subsubsection{\textbf{Access Delegation}}
 \begin{figure}[t]
  \centering
  \includegraphics[width=0.49\textwidth]{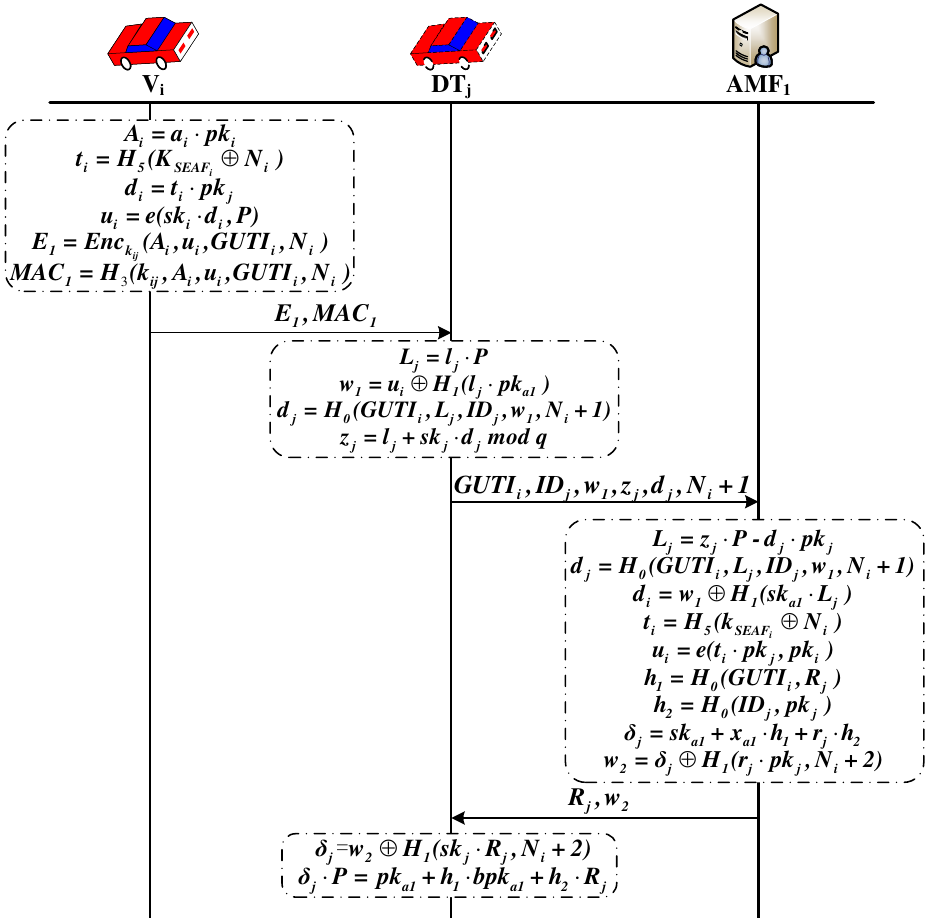}
  \caption{Access Delegation Phase}
  \label{fig:access}
\end{figure}
When attaching to the network for the first time, $MD_i$ and 5GC execute the standard 5G-AKA authentication protocol based on TS 33.501, and through the root key, sequentially calculate and derive with the current $AMF_1$ and $gNB_1$ the security anchor key $k_{SEAF_i}$, temporary session key $k_{gNB_1}$, and the globally unique temporary identifier $GUTI_i$ \cite{3gpp.36.331}. After that, $DT_j$ can apply the access delegation from $AMF_1$ for further access to the gNBs in the current domain as shown in Fig.~\ref{fig:access}.

\begin{enumerate}[Step 1:]
  \item $MD_i\rightarrow DT_j$: Authorized Token $(E_1,MAC_1)$\\  
     1a. $MD_i$ first randomly selects $a_i\in Z^*_q$ to compute $A_{i}=a_{i}\cdot pk_i$ for negotiating temporary session keys with the gNBs that may be accessed on subsequent trips.   Then, $MD_i$  generates the nonce number $N_i$ from the pseudo-random generator and uses the $k_{SEAF_i}$ to compute $t_{i}=H_5(k_{SEAF_i}\oplus N_i)$.\\
     1b. $MD_i$  uses the public key of $DT_j$ to compute $d_{i}=t_i \cdot pk_j$ and its private key   to compute $u_i=e(sk_i\cdot d_i, P)$ separately. \\
     1c. In order to prevent the disclosure and corruption, the $MD_i$ computes $E_1=Enc_ {k_{ij}}(A_{i},u_{i}, GUTI_i, N_i)$   and $MAC_1=H_3(k_{ij},A_{i},u_{i},GUTI_i,N_i)$ respectively.\\ 
     1d. Finally, $MD_i$ sends $(E_1,MAC_1)$ as the authorized token to $DT_j$, so that $DT_j$ can apply for the access delegation from $AMF_1$ in core network.
     \item $DT_j\rightarrow AMF_1$: Access Delegation Request $(GUTI_i, ID_{j},w_{1},z_{j},d_j,N_i+1)$\\
     2a.  Upon receiving the message from $MD_i$, $DT_j$ decrypts $E_1$ and verifies the correctness of $MAC_1$ and freshness of $N_i$ using the session key $k_{ij}$.  If verification fails, $DT_j$ discards the message and returns a failure response.  Otherwise, $DT_j$ randomly selects  $l_{j}\in Z_q^*$ to compute $L_{j}=l_{j}\cdot P$. \\
     2b. In order to protect $u_{i}$ against eavesdropping, $DT_j$ uses the public key of $AMF_1$ to computes $w_1=u_{i}\oplus H_1(l_{j}\cdot pk_{a1})$. \\
     2c. In order to prevent corruption, $DT_j$ computes $d_{j}=H_0(GUTI_i,L_{j},ID_{j},w_1,N_i+1)$ and uses its private key to compute $z_{j}=l_{j}+sk_{j}\cdot d_{j}\,mod\,q$.\\
     2d. Finally, $DT_j$ sends $(GUTI_i, ID_{j},w_{1},z_{j},d_j,N_i+1)$ to $AMF_1$ as access delegation request.

  \item $AMF_1\rightarrow DT_j$: Access Delegation Response  $(R_{j},w_2)$\\
     3a. Upon receiving request message from $DT_j$, $AMF_1$ first verifies the freshness of $N_i+1$. If so, $AMF_1$ then retrieves the corresponding public key $pk_j$ associated with $ID_j$ from its local database, computes $L_j' = z_j \cdot P - d_j \cdot pk_j$, and verifies the correctness of $d_j' {=} H_0(GUTI_i, L_j', ID_j, w_1, N_i+1)$. If correct, $AMF_1$ believes the request message is from legal $DT_j$ without attacks. Then, $AMF_1$ uses its private key to recover $u_i' = w_1 \oplus H_1(sk_{a1}\cdot L_j )$.\\
    3b. Based on the received $GUTI_i$, $AMF_1$ can search the corresponding $SUPI_i$ and anchor key $k_{SEAF_i}$ from local database with the help of $UDM$. $AMF_1$ then recomputes  $t_{i}'=H_5(k_{SEAF_i}\oplus N_i)$ and verifies the correctness of $u_i'=e(t_i'\cdot pk_j,pk_i)$. If the verification is failure, $AMF_1$ then aborts it. Otherwise, $AMF_1$ trusts that $DT_j$ has been authorized by legal $MD_i$ and further generates access delegation for $DT_j$ in order to prove its authority for gNBs. \\
    3c. Subsequently, $AMF_1$ calculates the access delegation $\delta_j$ for $DT_j$ as shown in equation \ref{equa1} where $r_j\in Z_q^*$ and $R_j=r_j\cdot P$. In order to prevent eavesdropping, $AMF_1$ computes $w_2=\delta_j\oplus H_2(r_{j}\cdot pk_{j}, N_i+2)$.\\
    3d.  Finally, $AMF_1$ transmits $(R_j,w_2)$ to $DT_j$ as access delegation response.
   \begin{equation}
  \begin{split}
&\delta_j=sk_{a1}+x_{a1}h_1+r_jh_2\,mod\,q\\
&h_1=H_0(GUTI_i,R_j)\\
&h_2=H_0(ID_j,pk_j)
\label{equa1}
\end{split}
\end{equation}
 \item 4a. Upon receiving message from $AMF_1$, $DT_j$ computes $N_i+2$ and utilizes its private key to decrypt  $ \delta_j=w_2\oplus H_2(sk_{j}\cdot R_j,N_i+2)$. \\
 4b. Then, $DT_j$ employs the public key pairs of $AMF_1$ to validate the correctness of $\delta_j\cdot P=h_1\cdot bpk_{a1}+h_2\cdot R_j+pk_{a1}$, where $bpk_{a1}=Y_{a1}+H_0(ID_{a1},pk_{a1})\cdot pk_{pub}$.  If verification fails, $DT_j$ aborts it and re-initiates the request. \\
 4c. If verification succeeds, $DT_j$ can accept $\delta_j$ as the valid access delegation and employ it to initiate the handover process with gNBs that the $MD_i$ will connect to within the current AMF domain.

 \end{enumerate}

\subsubsection{\textbf{Intra-AMF Handover}}

\begin{figure}[t]
  \centering
  \includegraphics[width=0.49\textwidth]{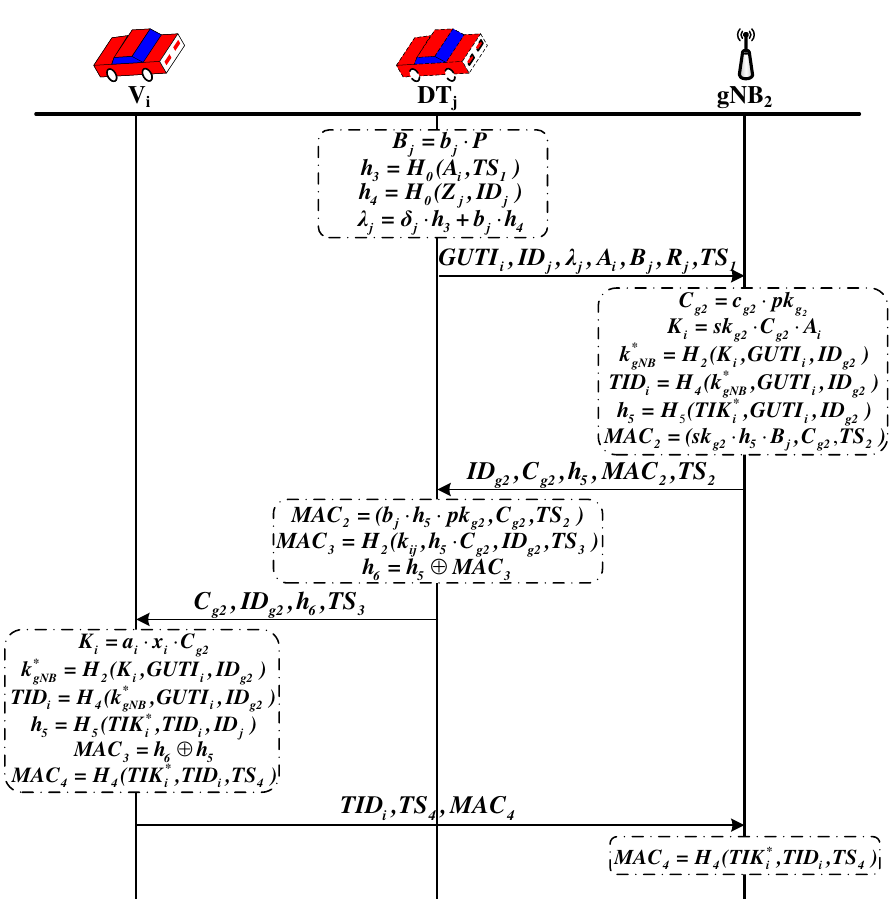}
  \caption{Intra-AMF Handover Authentication Phase}
  \label{fig:intra}
\end{figure}

When $MD_i$ is within the $gNB_1$'s coverage area, $DT_j$ predicts the target base station $gNB_2$ by analyzing $MD_i$'s movement trajectory in real-time. As shown in Fig.~\ref{fig:intra}, $DT_j$ initiates handover authentication with $gNB_2$ before $MD_i$ enters its coverage area, completing mutual authentication and session key negotiation between $MD_i$ and $gNB_2$ in advance.

 \begin{enumerate}[Step 1:]
   \item $DT_j\rightarrow gNB_2$: Handover Authentication Request: $(GUTI_i,ID_j,\lambda_j,A_i,B_j,R_j,TS_1)$\\
   1a. $DT_j$ randomly selects $b_j\in Z_q^*$ to compute $B_j=b_j\cdot P$. \\
   1b. In order to prevent corruption, $DT_j$ computes $h_3=H_0(A_i,TS_1)$ and $h_4=H_0(Z_j,ID_{j})$, where $TS_1$ is current timestamp and $Z_j=(sk_j+b_j)\cdot pk_{g2}$.\\
   1c. In order to prove the legitimacy of the identity, $DT_j$ computes  $\lambda_j=\delta_j\cdot h_3 +b_{j}\cdot h_4\, mod\, q$.\\ 
   1d. Finally, $DT_j$ transmits the handover authentication request $(GUTI_i,ID_j,\lambda_j,A_i,B_j,R_j,TS_1)$ to $gNB_2$ through the dedicated API channel.

   \item $gNB_2\rightarrow DT_j$: Handover Authentication Response $(ID_{g2},C_{g2},h_5,MAC_2,TS_2)$\\
  2a. Upon receiving the request message, $gNB_2$ first checks whether the freshness of the $TS_1'$ is within the time threshold $\triangle T$ in order to prevent replay attacks. If $|TS_1' - TS_1|>\triangle T$  exceeds the time threshold,  $gNB_2$ terminates the handover process. Otherwise, $gNB_2$ continues to check the legitimacy of $DT_j$ and whether it has been authorized by the $AMF_1$.\\
   2b. Based on $ID_j$, $gNB_2$ retrieves the corresponding public key $pk_j$ from its local database and subsequently recalculates $h_1'=H_0(GUTI_i',R_j')$, $h_2'=H_0(ID_j',pk_j)$, $h_3'=H_0(A_i',TS_1')$ and $h_4'=H_0(Z_j',ID_{j}')$ using the received parameters where $Z_j'=sk_{g2}\cdot (pk_j+B_j)$.\\
   2c. $gNB_2$ then verifies the correctness of $\lambda_j\cdot P\overset{\text{?}}{=}h_3'\cdot (pk_{a1}+h_1'\cdot bpk_{a1}+h_2'\cdot R_j)+h_4'\cdot B_j$. If the verification fails, $gNB_2$ aborts the handover process. Otherwise,  $gNB_2$ considers that $DT_j$ has been authorized by $AMF_1$. In addition, $gNB_2$ can optimize efficiency by performing batch verification as $\sum_j^n (\lambda_j) \cdot P\overset{\text{?}}{=}\sum_j^n (h_{3j}') \cdot pk_{a1}+\sum_j^n (h_{1j}'h_{3j}')\cdot bpk_{a1}+\sum_j^n (h_{3j}'h_{2j}')\cdot R_j+\sum_j^n h_{4j}'\cdot B_j$, when multiple handover requests are received.\\
   2d. $gNB_2$ selects $c_{g2}\in Z_q^*$ to compute $C_{g2}=c_{g2}\cdot sk_{g2}\cdot P$ and $K_i=sk_{g2}\cdot c_{g2}\cdot A_{i}$.  $gNB_2$ then generates the new temporary session key  $k_{gNB}^*=H_2(K_i,GUTI_i,ID_{g2})$  and further derives encryption key $TCK_i^*=KDF(k_{gNB}^*,\text{``Enc''})$ and integrity protection key $TIK_i^*=KDF(k_{gNB}^*, \text{``Int''})$ with $MD_i$  where $KDF$ is the key derivation function defined in 3GPP R.16 \cite{3gpp.36.331}.\\
   2e.  In order to prevent tracing on public wireless channel, $gNB_2$ locally generates temporary identification $TID_i=H_4(k_{gNB}^*,GUTI_i,ID_{g2})$.\\
  2f.  In order to prevent collusion, $gNB_2$ sequentially calculates $h_5=H_5(TIK_i^*,TID_i,ID_j)$ and $MAC_2=H_2(sk_{g2}\cdot h_5\cdot B_j,C_{g2},TS_2)$ where $TS_2$ is the current timestamp. \\
   2g.  Finally, $gNB_2$ transmits handover  response $(ID_{g2},C_{g2},h_5,MAC_2,TS_2)$ to $DT_j$ and retains the $(TID_i,k_{gNB}^*)$ in its local database.

   \item $DT_j\rightarrow MD_i$: Handover Authentication Notification $(C_{g2},ID_{g2},h_6,TS_3)$\\
   3a.  Upon receiving response message from $gNB_2$, the $DT_j$ first verifies the freshness of $TS_2$ to prevent replay attack. If $TS_2$ exceeds the time threshold, $DT_j$ aborts the handover process.  \\
   3b.  $DT_j$ utilizes the public key of $gNB_2$ to calculate $b_j\cdot h_5\cdot pk_{g2}$ and subsequently validates the correctness of $MAC_2$. Upon successful verification, $DT_j$ confirms the legitimacy of $gNB_2$ and the correctness of the received parameters.\\
   3c.  $DT_j$ generates the current timestamp $TS_3$ and computes $MAC_3 = H_2(k_{ij},  h_5\cdot C_{g2}, ID_{g2}, TS_3)$ and $h_6 = h_5 \oplus MAC_3$.\\
   3d.  Finally, $DT_j$ transmits $(C_{g2},ID_{g2},h_6,TS_3)$ to $MD_i$ as the  handover authentication notification via wireless channel.
   \item $MD_i\rightarrow gNB_2$: Handover Authentication Acknowledgment $(TID_i,MAC_4,TS_4)$\\
    4a. Upon receiving the notification message from $DT_j$, $MD_i$ first verifies the freshness of timestamp $TS_3$ to prevent replay attacks.\\
    4b. $MD_i$ computes the shared secret value $K_i = a_i\cdot x_i \cdot C_{g2}$ and subsequently generates the temporary session key $k_{gNB}^* = H_2(K_i, GUTI_i, ID_{g2})$ with $gNB_2$. \\
    4c. $MD_i$ utilizes $k_{gNB}^*$ to compute the temporary identity identifier $TID_i = H_4(k_{gNB}^*, GUTI_i, ID_{g2})$, and subsequently derives the temporary encryption key $TCK_i^*=KDF(k_{gNB}^*,\text{``Enc''})$ and temporary integrity protection key $TIK_i^*=KDF(k_{gNB}^*,\text{``Int''})$ with $gNB_2$.\\
    4d. $MD_i$ computes $h_5' = H_5(TIK_i^*, TID_i, ID_j)$ and recovers $MAC_3'=h_6\oplus h_5'$. Subsequently, $MD_i$ employs  $k_{ij}$ with $DT_j$ to verify whether $MAC_3'\stackrel{?}{=}H_2 (k_{ij}, h_5\cdot C_{g2}, ID_{g2},  TS_3)$ is valid.\\
    4e. Upon successful verification, when $MD_i$ enters the signal coverage area of $gNB_2$, it generates current timestamp $TS_4$ and computes $MAC_4 = H_4(TIK_i^*, TID_i, TS_4)$. Then, $MD_i$ transmits $(TID_i, MAC_4, TS_4)$ to $gNB_2$ as the handover authentication acknowledgment message through the open wireless channel.\\
    4f. If any of the aforementioned verification steps fails, upon entering the coverage area of $gNB_2$, $MD_i$ will re-execute the standard handover authentication procedure according to the 5G-AKA protocol defined in TS 33.501  to ensure the establishment of a secure access connection with $gNB_2$ \cite{3gpp.36.331}.
   \item 5a. Upon receiving the acknowledgment message from $MD_i$, $gNB_2$ first checks the freshness of $TS_4$. If so, $gNB_2$ retrieves the $k_{gNB}^*$ from the database based on the $TID_i$ and further verifies the validity of $MAC_4=H_2(TIK_i^*,TID_i,TS_4)$.\\
   5b. After all verifications pass, $MD_i$ and $gNB_2$ complete the handover authentication protocol and establish a secure wireless communication link.
 \end{enumerate}

 \subsubsection{\textbf{Inter-AMF Handover}}
\begin{figure}[t]
  \centering
  \includegraphics[width=0.49\textwidth]{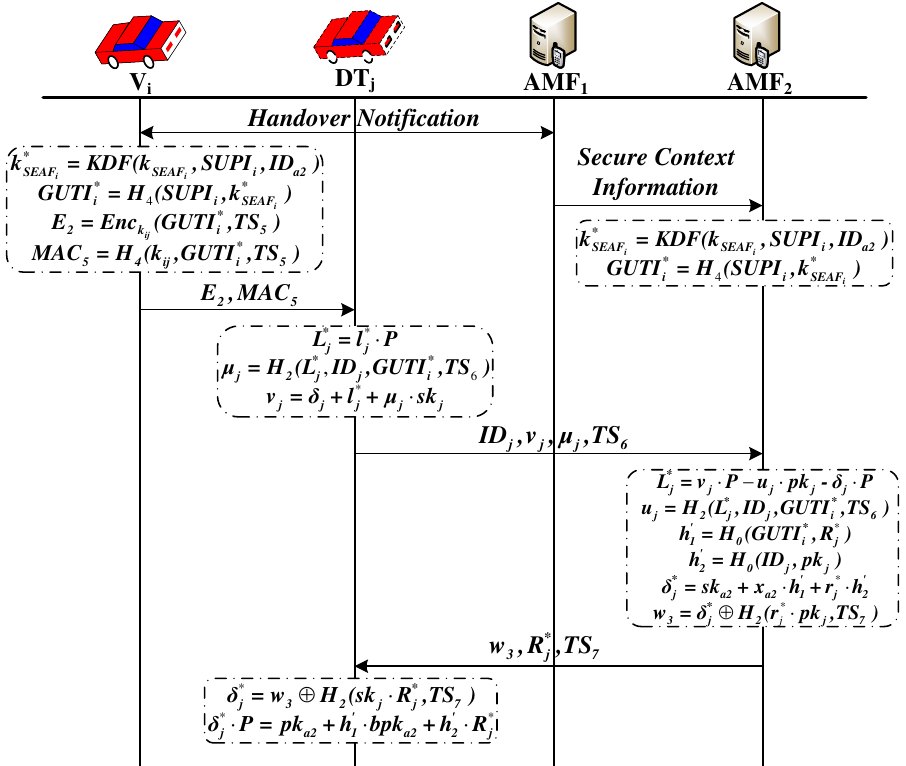}
  \caption{Inter-AMF Handover Authentication Phase}
  \label{fig:inter}
\end{figure}

 When $MD_i$ is about to move from $gNB_2$ under $AMF_1$ domain to $gNB_3$ under $AMF_2$ domain, based on the analysis of the movement trajectory, $DT_j$ notifies the $AMF_1$ and $MD_i$ to trigger inter-AMF handover authentication, ensuring seamless handover authentication for $MD_i$ as shown in Fig~\ref{fig:inter}. The details are as follows.

 \begin{enumerate}[Step 1:]
 \item $AMF_1\rightarrow AMF_2$:  Security Context Information $(SUPI_i,k_{SEAF_i},ID_j,\delta_j\cdot P)$\\
    1a. To prepare for the inter-AMF handover, $AMF_1$ transmits the security context information of $MD_i$ including $(SUPI_i,k_{SEAF_i},ID_j,\delta_j\cdot P)$ to $AMF_2$ through N14 interface.\\
    1b. $AMF_2$ then generates the new anchor key $k_{SEAF_i}^*=KDF(k_{SEAF_i},SUPI_i,ID_{a2})$ and new globally unique temporary identity $GUTI_i^*=H_4(SUPI_i,k_{SEAF_i}^*)$ with $MD_i$. In addition, $AMF_2$ stores $(GUTI_i^*,ID_j,k_{SEAF_i}^*,\delta_j\cdot P)$ in the local database.
    
\item $MD_i\rightarrow DT_j$: $(E_2,MAC_5)$\\
    2a. Upon receiving notification from $DT_j$, $MD_i$ first updates the new anchor key $k_{SEAF_i}^*=KDF(k_{SEAF_i},SUPI_i,ID_{a2})$ and corresponding globally unique temporary identity $GUTI_i^*=H_4(SUPI_i,k_{SEAF_i}^*)$.\\
    2b. $MD_i$ employs the session key $k_{ij}$ to calculates $E_2=Enc_{k_{ij}}(GUTI_i^*,TS_5)$ and  authentication code $MAC_5=H_4(k_{ij},GUTI_i^*,TS_5)$ where $TS_5$ is the current timestamp. \\
    2c. Finally, $MD_i$ transmits  $(E_2,MAC_5)$ to $DT_j$.

\item $DT_j\rightarrow AMF_2$: Access Delegation Request $(ID_j,v_j,\mu_j,TS_6)$\\
    3a. Upon receiving $(E_2,MAC_5)$, $DT_j$ first verifies the correctness of $MAC_5$ and freshness of $TS_5$ after decrypting $E_2$. If verifications are correct, $DT_j$ then randomly selects $l_{j}^*\in Z_q^*$ to compute $L_{j}^*=l_{j}^*\cdot P$. \\
    3b.  In order to prevent collusion, $DT_j$ computes $\mu_j=H_2(L_j^*,ID_j, GUTI_i^*,TS_6)$ and $v_j=\delta_j + l_j^*+\mu_j\cdot sk_j  \,mod\,q$,  where $TS_6$ is current timestamp.\\
    3c. $DT_j$ submits $(ID_j, v_j, \mu_j, TS_6)$ to $AMF_2$ for requesting a new access delegation within the current domain.
 \item $AMF_2\rightarrow DT_j$: Access Delegation Response $(w_3,R_{j}^*,TS_7)$\\
    4a. Upon receiving the request message from $DT_j$, $AMF_2$ first verifies the freshness of $TS_6$. Then, $AMF_2$ searches $(GUTI_i^*,ID_j,\delta_j\cdot P)$ from database. \\
    4b. $AMF_2$ then  computes $L_{j}^*\overset{\text{?}}{=}v_j\cdot P-\delta_j\cdot P-\mu_j'\cdot pk_j$ and verifies whether $\mu_j'=H_2(L_j^*,ID_j, GUTI_i^*,TS_6)$ is correct. If so, $AMF_2$ then can generate the new access delegation $\delta_j^*$ for $DT_j$ to access the gNBs in the current domain as equation 1 in Section 4.3.3. \\
    4c. $AMF_2$ computes $w_3=\delta_j^*\oplus H_2(r_{j}^*\cdot pk_{j},TS_7)$ and sends $(w_3,R_{j}^*,TS_7)$ to the $DT_j$ where $R_{j}^*=r_{j}^*\cdot P$ and $r_{j}^*\in Z_q^*$, and $TS_7$ is current timestamp.
\item Upon receiving the message, $DT_j$ first verifies the freshness of $TS_7$. If so, $DT_j$  then employs its private key $sk_j$ to decrypt $\delta_j^*$ from $w_3$ and verifies the correctness of $\delta_j^*$. If so, $DT_j$ will initiate the handover procedure with the $gNB_3$ that the $MD_i$ is about to connect to, as described in section 4.3.4, to assist in the mutual authentication and key negotiation between the $MD_i$ and $gNB$  that will be accessed in advance.
 \end{enumerate}

\section{Security Analysis}
In this section, we evaluate the security of the proposed scheme through formal and informal analysis. First, we use the BAN-logic, RoR model, and ProVerif to demonstrate the mutual authentication, key agreement, and session key security between MD and gNB. In addition, we also employ the informal security analysis to further analyze and prove that the proposed scheme can achieve the security goals defined in Section 3.3.

\subsection {Formal Verification based on BAN-Logic}
BAN logic is model logic based on subject knowledge and belief reasoning. We first list the common rule of BAN logic that will be used in the proof as follows:

\begin{enumerate}
  \item The fresh-promotion rule: $\frac{{P}|{\equiv \sharp(X)}}{{P}|{\equiv \sharp(X,Y)}}$
  \item The nonce-verification rule:  $\frac{{P}|{\equiv \sharp(X)},{P}|{\equiv Q}|{\sim X}}{{P}|{\equiv Q}|{\equiv X}}$
  \item The decomposition rule: $\frac{{P}|{\equiv Q}|{\equiv (X,Y)}}{{P}|{\equiv Q}|{\equiv X}},\frac{{P}|{\equiv (X,Y)}}{{P}|{\equiv X}}$
  \item The composition rule: $\frac{{P}|{\equiv X},{P}|{\equiv Y}}{{P}|{\equiv (X,Y)}}$
  \item The jurisdiction rule:  $\frac{{P}|{\equiv Q}|{\Rightarrow X},{P}|{\equiv Q}|{\equiv X}}{{P}|{\equiv X}}$
  \item The message-meaning rule: $\frac{{P}|{ \equiv P \stackrel{K}\longleftrightarrow Q},P\{X\}_K}{{P}|{\equiv Q}|{\sim X}}$
\end{enumerate}

 Then, the basic goal of the proposed scheme is to achieve the mutual authentication and key negotiation between $MD_i$ and $gNB_2$ in BAN-logic is listed as follows:
\begin{enumerate}
  \item Goal 1: ${MD_i}$$|$${{ \equiv gNB_2} \stackrel{k_{gNB}^*}\longleftrightarrow { MD_i}}$
  \item Goal 2: ${ gNB_2}$$|$${ {\equiv gNB_2} \stackrel{k_{gNB}^*}\longleftrightarrow { MD_i}}$
  \item Goal 3: ${ MD_i}$$|$${\equiv gNB_2}$$|$${ {\equiv gNB_2 }\stackrel{ k_{gNB}^*}\longleftrightarrow { MD_i}}$
  \item Goal 4: ${ gNB_2}$$|$${ \equiv MD_i}$$|$${ {\equiv MD_i} \stackrel{ k_{gNB}^*}\longleftrightarrow { gNB_2}}$
\end{enumerate}

In addition, we give the necessary assumptions as follows in order to better analyze the proposed scheme:
\begin{enumerate}
  \item Assumption 1: ${MD_i}$$|$${{ \equiv MD_i} \stackrel{k_{ij}}\longleftrightarrow { DT_j}}$
  \item Assumption 2: ${DT_j}$$|$${{ \equiv MD_i} \stackrel{k_{ij}}\longleftrightarrow { DT_j}}$
  \item Assumption 3: ${gNB_2}$$|$${\equiv DT_j}$$|$$\Rightarrow$$(GUTI_i,ID_j,\lambda_j,A_i,B_j,$\\$R_j,TS_1)$
  \item Assumption 4: ${DT_j}$$|$${\equiv gNB_2}$$|$$\Rightarrow$$(ID_{g2},C_{g2},h_5,MAC_2,$\\$TS_2)$
  \item Assumption 5: $MD_i$$|$${\equiv DT_j}$$|$$\Rightarrow$$(C_{g2},ID_{g2},h_6,TS_3)$
\end{enumerate}

  Now, we prove that the proposed scheme can achieve from Goal 1 to Goal 4 as follows.
  
  Since $gNB_2$ receives the message $(GUTI_i,ID_j,\lambda_j,A_i,$\\$B_j,R_j,TS_1)$, we have:

  S1: $gNB_2$ $\triangleleft$ $(GUTI_i,ID_j,\lambda_j,A_i,B_j,R_j,TS_1)$

 In the proposed scheme, $gNB_2$ verifies the freshness of $TS_1$ to prevent replay attacks. Once verification is correct, we have:

 S2: $gNB_2$ $|$${ \equiv}$ $\sharp (TS_1)$

  According to the fresh-promotion rule and S2, we have:

S3: $gNB_2$ $|$${ \equiv}$ $\sharp (GUTI_i,ID_j,\lambda_j,A_i,B_j,R_j,TS_1)$

 In the proposed scheme, $gNB_2$ verifies the correctness of $\lambda_j P\overset{\text{?}}{=}h_3'(pk_{a1}+h_1'bpk_{a1}+h_2'R_j)+h_4'B_j$. If so, according to the S3, we have:

 S4: $gNB_2$ $|$${ \equiv}$ $ DT_j$ $|$$\sim$  $ (GUTI_i,ID_j,\lambda_j,A_i,B_j,R_j,TS_1)$

 According to the S3, S4 and nonce-verification rule, we have:

 S5: $gNB_2$ $|$${ \equiv}$ $ DT_j$ $|$${ \equiv}$  $ (GUTI_i,ID_j,\lambda_j,A_i,B_j,R_j,TS_1)$

According to the S5, Assumption 3 and the justification rule, we have:

S6: $gNB_2$ $|$${ \equiv}$   $ (GUTI_i,ID_j,\lambda_j,A_i,B_j,R_j,TS_1)$

According to the S6 and decomposition rule, we have:

S7: $gNB_2$  $|$${ \equiv}$  $A_i$

And,

S8: $gNB_2$  $|$${ \equiv}$  $GUTI_i$

In the proposed scheme, $gNB_2$ has the private key  $sk_{g2}$, we have:

S9: $gNB_2$ $|$${ \equiv}$  $sk_{g2}$

Due to the $gNB_2$ randomly selects $c_{g2}\in Z_q^*$, we have

S10: $gNB_2$ $|$${ \equiv}$  $c_{g2}$

In the proposed scheme, $gNB_2$ computes the secret value $K_i=c_{g2}\cdot sk_{g2}\cdot A_{i}$. According to the S7, S9, S10, and the composition rule, we have:

S11: $gNB_2$ $|$${ \equiv}$  $K_i$.

Since $ID_{g2}$ is the identification  of $gNB_2$, we have:

S12: $gNB_2$ $|$${ \equiv}$ $ID_{g2}$.

Since $k_{gNB}^*=H_2(K_i,GUTI_i,ID_{g2})$, according to the S8, S11, S12 and the composition rule, we have: 

S13: $gNB_2$ $|$${ \equiv}$  $k_{gNB}^*$.

That is, ${ gNB_2}$$|$${ {\equiv gNB_2} \stackrel{k_{gNB}^*}\longleftrightarrow { MD_i}}$ $\qquad\qquad\quad$ (Goal 2)

Since $DT_j$ receives the $(ID_{g2},C_{g2},h_5,MAC_2,TS_2)$ from $gNB_2$, we have:

S14: $DT_j$ $\triangleleft$ $(ID_{g2},C_{g2},h_5,MAC_2,TS_2)$

$DT_j$ checks the freshness of $TS_2$ to defend replay attack. If so, we have:
 
S15: $DT_j$ $|$${ \equiv}$ $\sharp (TS_2)$

According to the fresh-promotion rule and S15, we have:

S16: $DT_j$ $|$${ \equiv}$ $\sharp (ID_{g2},C_{g2},h_5,MAC_2,TS_2)$ 

In the proposed scheme, $DT_j$ uses $b_j$ and $pk_{g2}$ to compute  $MAC_2'=H_2(b_j\cdot h_5\cdot pk_{g2},C_{g2},TS_2)$. Then, $DT_j$ verifies the correctness of $MAC_2'=MAC_2$. If so, we have:

S17: $DT_j$ $|$${ \equiv}$ $gNB_2$ $|$$\sim (ID_{g2},C_{g2},h_5,MAC_2,TS_2)$ 

According to the S16, S17 and the nonce-verification rule, we have:

S18: $DT_j$ $|$${ \equiv}$ $gNB_2$ $|$$\equiv  (ID_{g2},C_{g2},h_5,MAC_2,TS_2)$

According to the S18, Assumption 4 and the jurisdiction rule, we have:

S19: $DT_j$ $|$${ \equiv}$  $  (ID_{g2},C_{g2},h_5,MAC_2,TS_2)$

According to the S19 and the decomposition rule, we have:

S20: $DT_j$ $|$${\equiv}$ $ (ID_{g2},C_{g2},h_5)$ 

After receiving the  $(C_{g2},ID_{g2},h_6,TS_3)$, $MD_i$ first verifies the freshness of $TS_3$ to prevent replay attack. If so, we have:

S21: $MD_i$ $|$${ \equiv}$ $\sharp (TS_3)$

According to the fresh-promotion rule and S20, we have:

S22: $MD_i$ $|$${ \equiv}$ $\sharp (C_{g2},ID_{g2},h_6,TS_3)$ 

$MD_i$ first computes $h_5'$, and then uses $k_{ij}$ to verify the correctness of $MAC_3'$. Since $MAC_3 = H_2(k_{ij}, h_5\cdot C_{g2}, ID_{g2}, TS_3)$ and $h_6= h_5\oplus MAC_3$. If the verification is correct, we have:

S23: $MD_i$ $|$${ \equiv}$ $DT_j$ $|$$\sim (C_{g2},ID_{g2},h_5,TS_3)$

Since $MAC_3$ is composed of $h_5$ and $TS_3$, according to the S21, we have:

S24: $MD_i$ $|$${ \equiv}$ $\sharp (C_{g2},ID_{g2},h_5,TS_3)$ 

According to the S23 and S24, we have:

S25: $MD_i$ $|$${ \equiv}$ $DT_j$ $|$${ \equiv}(C_{g2},ID_{g2},h_5,TS_3)$

According to the decomposition rule and S25, we have:

S26: $MD_i$$|$${ \equiv}(C_{g2},ID_{g2},h_5,TS_3)$

According to the decomposition rule and S26, we have:

S27: $MD_i$$|$${ \equiv} C_{g2}$

And, 

S28: $MD_i \mid \equiv ID_{g2}$

Since $MD_i$ has the private key $sk_i$, we have:

S29: $MD_i$ $|$${ \equiv}$ $sk_i$

Due to the $MD_i$ randomly selects $a_{i}\in Z_q^*$, we have

S30: $MD_i$ $|$${ \equiv}$  $a_{i}$

In the proposed scheme, $MD_i$ computes the $ K_i= a_i\cdot sk_i \cdot C_{g2}$. According to the S27, S29, S30 and the composition rule, we have:

S31: $MD_i$ $|$${ \equiv}$  $K_i$

Since $GUTI_i$ is the temporary identification  of $MD_i$, we have:

S32: $MD_i$ $|$${ \equiv}$ $GUTI_i$

Since $k_{gNB}^*=H_2(K_i,GUTI_i,ID_{g2})$, according to the S28, S31, S32 and the composition rule, we have:

S33: $MD_i$ $|$${ \equiv}$ $k_{gNB}^*$

That is, ${ MD_i}$$|$${ {\equiv gNB_2} \stackrel{k_{gNB}^*}\longleftrightarrow { MD_i}}$ $\qquad\qquad\quad\quad$ (Goal 1)























 $gNB_2$ receives the $(GUTI_i,MAC_4,TS_4)$ from $MD_i$ where $MAC_4=H_2(TIK_i^*,TID_i,TS_4)$, since $TIK_i^*$ is derived by $k_{gNB}^*$, according to the S13 and the message-meaning rule, we have:

 S34:  $gNB_2$ $|$${ \equiv}$ $MD_i$ $|$$\sim$  $ (k_{gNB}^*,TID_i,TS_4)$

 $gNB_2$ can check the freshness of $TS_4$ to prevent replay attack. If so, we have:

S35: $gNB_2$ $|$${ \equiv}$ $\sharp (TS_4)$

According to the S35 and  the fresh-promotion rule, we have:

S36:  $gNB_2$ $|$${ \equiv}$ $\sharp (k_{gNB}^*,TID_i,TS_4)$

According to the S34, S36 and the nonce-verification rule, we have:

S37: $gNB_2$ $|$${ \equiv}$ $MD_i$  $|$${ \equiv}$ $(k_{gNB}^*,TID_i,TS_4)$

According to the S37 and the decomposition rule, we have:

S38: $gNB_2$ $|$${ \equiv}$ $MD_i$  $|$${ \equiv}$ $k_{gNB}^*$

 That is,   ${ gNB_2}$$|$${ \equiv MD_i}$$|$${ {\equiv MD_i} \stackrel{ k_{gNB}^*}\longleftrightarrow { gNB_2}}$ $\quad$  $\,$  (Goal 4)

$MD_i$ can calculate $h_5$ using $TIK_i^*$ derived by $k_{gNB}^*$, and verify its correctness by checking whether $MAC_3$ is valid. If so, we have:

 S39:  $MD_i$ $|$${ \equiv}$ $gNB_2$ $|$$\sim$  $ h_5$

  According to the  S24, and the fresh-promotion rule, we have:

  S40:  $MD_i$ $|$${ \equiv}$ $\sharp (h_5)$

  According to the S39, S40  and the nonce-verification rule, we have:

  S41:  $MD_i$ $|$${ \equiv}$ $gNB_2$  $|$${ \equiv}$ $ h_5$

  Since $h_5=H_2(TIK_i^*,TID_i,ID_j)$, according to the S40 and the decomposition rule, we have:

  S42: $MD_i$ $|$${ \equiv}$ $gNB_2$  $|$${ \equiv}$ ${TIK_i^*}$

  Since $TIK_i^*$ is directly derived by $k_{gNB}^*$. That is, ${ MD_i}$$|$${\equiv gNB_2}$$|$${ {\equiv gNB_2 }\stackrel{ k_{gNB}^*}\longleftrightarrow { MD_i}}$ $\qquad\qquad\quad  \quad$ (Goal 3)

\subsection{Formal Analysis based on RoR Model}
To validate the session key security of our proposed scheme, we then perform a formal security analysis based on the Real-or-Random (RoR) model \cite{abdalla2005password}.

\subsubsection{Participants}In the proposed scheme, three participating entities are involved in the handover authentication process: $MD_i$, $DT_j$, and $gNB_2$. Let $\Pi_{MD_i}^m$, $\Pi_{DT_j}^d$, and $\Pi_{gNB_2}^g$ denote instances $m$, $d$, and $g$ of $MD_i$, $DT_j$, and $gNB_2$, respectively.
\subsubsection{Partnering}Two instances are considered partners if and only if they simultaneously satisfy the following three conditions: (1) both instances have reached an accepted state; (2) the instances have mutually authenticated each other and share an identical session identifier; and (3) the instances have established a mutual partnership.
\subsubsection{Freshness} An instance is considered fresh if its session key established with another instance has not been revealed to adversary $\mathcal{A}$.
\subsubsection{Adversary} Adversary $\mathcal{A}$ has the capability to read, modify, delete, and fabricate messages transmitted over the public communication channel. In addition, $\mathcal{A}$ has the attack capabilities according to the following queries: 
\begin{itemize}
    \item $Execute(\Pi_{MD_i}^m, \Pi_{DT_j}^d,\Pi_{gNB}^g)$: The $\mathcal{A}$ can obtain all messages exchanged and transmitted between $MD_i$, $DT_j$ and $gNB$ by executing this query.
    
    \item $Send(\Pi_{MD_i}^m, \Pi_{DT_j}^d,\Pi_{gNB}^g,m)$:  The $\mathcal{A}$ can launch the active attack and send the message $m$ to the $ \Pi_{MD_i}^m$ and $\Pi_{gNB}^g$. If the message $m$ passes the verification from the instance, $\mathcal{A}$  will receive a valid response. Otherwise, the query will abort.  
    
    \item $Reveal(\Pi^t)$: In this query, the $\mathcal{A}$ can obtain the current specific state information created by the $\Pi_{MD_i}^m$ and its partner $\Pi_{gNB}^g$ in the current session.
    
    \item $Corrupt(\Pi^t)$: Upon executing this query,  the instances $\Pi_{MD_i}^m$ and  $\Pi_{gNB}^g$ can be compromised,  and its long-term secret keys can be leaked to the $\mathcal{A}$.
    
    \item $Test(\Pi_{MD_i}^m,b)$: This query verifies the semantic security of the temporary session key between $\Pi_{MD_i}^t$ and $\Pi_{gNB}^u$. The output is determined by flipping a uniformly random coin $b$. For an established and fresh session key, the query returns the real session key if $b=1$, and a random string of equal length if $b=0$. The  $\mathcal{A}$'s goal is to guess the hidden bit $b$. If $\mathcal{A}$ can guess $b$ correctly, it breaks the semantic security of the session key. 
\end{itemize}

In the proposed scheme, the cryptographic hash function $h(\cdot)$ is accessible to all   participants and adversary 
 $\mathcal{A}$, and is modeled as a random oracle $\mathcal{HO}$, where hash query returns a fixed but unpredictable value for each unique input.

 \subsubsection{Semantic security} Let $Succ$ be the event that $\mathcal{A}$ win the game, and $Pr[Succ]$ denote its probability.  We use the $Adv^P_\mathcal{A}=|2Pr[Succ]-1|$ to represent the advantage of $\mathcal{A}$ that can break the semantic security of the proposed scheme. If the advantage $Adv^P_\mathcal{A}$ is bounded by a negligible value $\epsilon$, we consider the proposed scheme secure against polynomial-time adversaries.
 
\textit{Theorem 1}:  Let  $\mathcal{A}$  be a t-polynomial time adversary against the semantic security of our proposed scheme.  $q_s$, $q_e$ and $q_h$ represent the number of
			Send queries, Execute queries and Hash queries respectively. $|Hash| $
			denotes  the bit lengths
			of the hash function.  $Adv_{\mathcal{A}}^{ECDHP}$  is the advantage of
			adversary in breaking the ECDH problem in upper-bound time $t$. The advantage of $\mathcal{A}$ in breaching the session key security of our proposed scheme is  estimated as:			
				\begin{equation}
	Adv_{\mathcal{A}}^P \leq \frac{q_h^2}{|Hash|}+\frac{(q_s+q_e)^2}{|Hash|}+2q_hAdv_{\mathcal{A}}^{ECDHP}
			\end{equation}
            
 \textit{Proof:} We define the four games $Game_i$ to verify the semantic security of the proposed scheme, where $i=[0,3]$.

 $Game_0$: This game directly simulates the real attack scenario where  $\mathcal{A}$ attempts to determine the hidden bit $b$ chosen at the game's initialization. According to the semantic security of the session key, we have:
    \begin{equation}
   Adv^P_\mathcal{A}=|2Pr[Succ_0]-1|
       \label{eq:1}
   \end{equation}

 $Game_1$: This game represents an eavesdropping attack where the $\mathcal{A}$ can intercept messages transmitted between $DT_j$, $MD_i$ and $gNB$ over the public channel by executing $Execute$ query. The $\mathcal{A}$ subsequently uses the $Test$ query to distinguish whether the output key is real or random. The temporary session key between $MD_i$ and $gNB_2$ is computed as $k_{gNB}^*=H_2(K_i,GUTI_i,ID_{g2})$, where $K_i=a_i\cdot sk_i\cdot C_{g2}=c_{g2}\cdot sk_{g2}\cdot A_i$. To obtain the $k_{gNB}^*$, the $\mathcal{A}$ would need to acquire the secret values $a_i$ and $c_{g2}$ as well as the private keys  $sk_i$ and $sk_{g2}$. However, these values cannot be extracted from the messages captured over the channel. That is, the adversary's advantage in winning $Game_1$ does not increase. Therefore, we have:
  \begin{equation}
Pr[Succ_1]=Pr[Succ_0]
       \label{eq:1}
   \end{equation}

$Game_2$: Based on $Game_1$, $Game_2$ adds send query and hash query. This game simulates that the adversary $\mathcal{A}$ can launch active attacks through forging messages that can be accepted and verified by receivers. $\mathcal{A}$ can intercept all exchanged messages $<GUTI_i, ID_j,\lambda_j,A_i,B_j,R_j,TS_1>$, $<ID_{g2},C_{g2},h_5,MAC_2,TS_2>$, $<C_{g2},ID_{g2},h_6,TS_3>$ and $<TID_i,MAC_4,TS_4>$ to continuously execute hash query to find the message collisions. In addition, $\mathcal{A}$ can embed the timestamp from intercepted messages into forged messages and search for collisions through hash query. According to the birthday paradox, we have:
   \begin{equation}
|Pr[Succ_2]-Pr[Succ_1]|\leq \frac{q_h^2+(q_e+q_s)^2}{2|Hash|}
       \label{eq:1}
   \end{equation}


$Game_3$: In this game, the adversary can launch  the $Corrupt(\Pi^t)$  and $Reveal(\Pi^t)$ to compromise the temporary session key $k_{gNB}^*$ between $gNB$ and $MD_i$. $\mathcal{A}$ can execute $Exectute$, $Hash$ queries, while leveraging combinations of $Reveal$ and $Corrupt$ queries. This game occurs on three occasions:
\begin{itemize}
    \item $Reveal(\Pi^t)$: This query is to prove the ephemeral secret leakage. By executing this query, the $\mathcal{A}$ can obtain the state information $(a_i,A_i)$ of the instance  $\Pi^m_{MD_i}$ and $(c_{g2},C_{g2})$ of the instance  $\Pi^g_{gNB}$. To compute the temporary session key $k_{gNB}^*$, the $\mathcal{A}$ needs to acquire both $sk_i$ and $sk_{g2}$.
    
    \item  $Corrupt(\Pi^{t}):$ This query is to prove the perfect forward secrecy. By executing this query, the $\mathcal{A}$ can obtain the $(sk_i,A_i)$ of the instance  $\Pi^m_{MD_i}$ and   $(sk_{g2},C_{g2})$ of the instance  $\Pi^g_{gNB}$.  To compute the temporary session key $k_{gNB}^*$, the $\mathcal{A}$ needs to acquire both $a_i$ and $c_{g2}$.

    \item $Corrupt(\Pi^{t})$ and $Reveal(\Pi^{t})$: By executing these queries, the adversary $\mathcal{A}$ can obtain either the secret key $sk_i$ from instance $\Pi^m_{MD_i}$ and the state information $(c_{g2},C_{g2})$ from instance $\Pi^g_{gNB}$, or the state information $(a_i,A_i)$ from instance $\Pi^m_{MD_i}$ and the secret key $sk_{g2}$ from instance $\Pi^g_{gNB}$. To compute the temporary session key $k_{gNB}^*$, $\mathcal{A}$ must obtain either the pair $(a_i, sk_{g2})$ or the pair $(c_{g2}, sk_i)$.
\end{itemize}

The probability for $\mathcal{A}$ to distinguish between $Game_2$ and $Game_3$ and derive the shared session key $k_{gNB}^*$ is negligible without breaking the hash function or solving the ECDHP. Therefore, we have:
   \begin{equation}
|Pr[Succ_3]-Pr[Succ_2]|\leq q_h Adv_\mathcal{A}^{ECDHP}
       \label{eq:1}
   \end{equation}



After exhausting all query options and implementing various attacks against the proposed scheme,  the adversary $\mathcal{A}$ can only guess the bit $b$ to win the game through the $Test(\Pi^m_{MD_i},b)$ query. Therefore, we have:
   \begin{equation}
   Pr[Succ_3]|=\frac{1}{2}
       \label{eq:1}
   \end{equation}

According to the the triangle inequality $ |a \pm b| \leq |a| + |b| $, we have:
\begin{equation}
\begin{split}
\frac{1}{2}Adv_{\mathcal{A}}^P &= \left|Pr[Succ_0]-\frac{1}{2}\right| \\
                            &= \left|Pr[Succ_0]-Pr[Succ_3]\right| \\
                            &\leq \frac{q_h^2}{2|Hash|}+\frac{(q_s+q_e)^2}{2|Hash|}+q_hAdv_{\mathcal{A}}^{ECDHP}
\end{split}
\end{equation}

That is $Adv_{\mathcal{A}}^P \leq \frac{q_h^2}{|Hash|}+\frac{(q_s+q_e)^2}{|Hash|}+2q_hAdv_{\mathcal{A}}^{ECDHP}$. Given that $|Hash| = 2^q$, where $q$ represents the output bit length of the hash function, and $Adv_{\mathcal{A}}^{ECDHP}$ denotes the probability of successfully solving the ECDH problem (which is negligibly small under reasonable security assumptions), we can conclude that $Adv_{\mathcal{A}}$ is negligible. Consequently, the proposed scheme achieves semantic security of session keys under the RoR model.

\subsection{Formal Analysis based on ProVerif}
ProVerif is an automated cryptographic protocol verification tool widely used to examine whether protocols satisfy specific security properties based on the Dolev-Yao attack model \cite{blanchet20212}. We continue to use ProVerif to demonstrate that the proposed scheme can achieve mutual authentication, key agreement, and data confidentiality between gNB and MD with the assistance of DT during the handover. For space conservation, the complete verification code for the proposed scheme is presented in \cite{GuanjieLi_DTHA_2025}.

\begin{figure}[tbp]
    \centering
    \subfloat[Events and Queries]{
        \includegraphics[width=0.38 \textwidth]{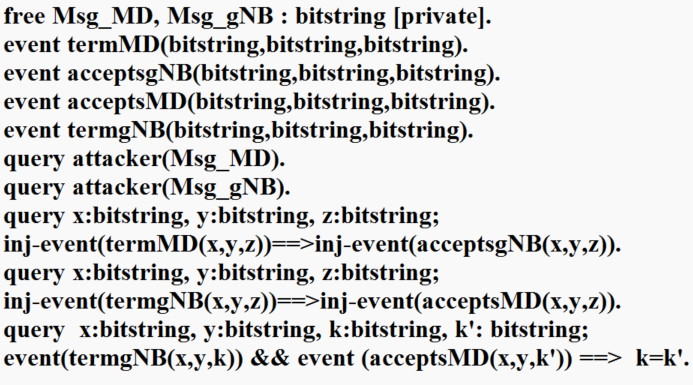}
    }\\
    \subfloat[Simulation Results]{
        \includegraphics[width=0.38\textwidth]{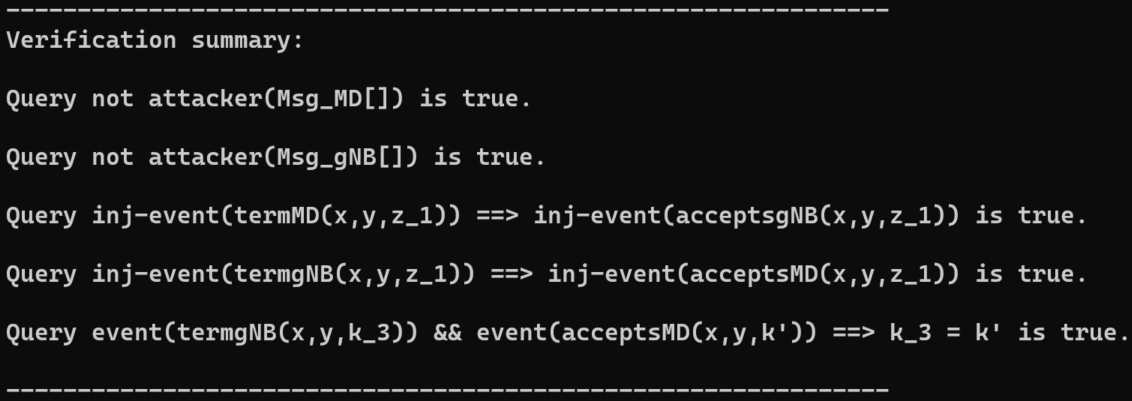}
    }
 
    \caption{Queries and Results in ProVerif}
    \label{query}
\end{figure}

In order to verify that the proposed scheme can achieve the security objectives, we have declared the necessary events and defined corresponding queries as shown in Fig~\ref{query}(a). $query \; attacker(Msg\_MD)$ and $query \; attacker(Msg\_gNB)$ describe the confidentiality of transmitted messages encrypted by the negotiated session key between gNB and MD, $inj-event(termMD(x,y,z))==>inj-event$$(acceptsgNB(x,y,z))$ describes the authentication of MD to gNB,  $inj-event(termgNB(x,y,z))==>inj-event(acceptsMD(x,y,z))$ describes the authentication of gNB to MD similarly. Furthermore, the last query $event(termgNB(x,y,k)) \&\& event (acceptsMD(x,y,k'))$\quad $==>  k=k'$ describes that gNB and MD can negotiate the same session key after completing mutual authentication. Fig.~\ref{query}(b) presents the simulation results of the proposed scheme under ProVerif, it demonstrates that the proposed scheme has successfully achieved mutual authentication, key negotiation as well as data confidentiality.

\subsection{Further Security Analysis}
We further demonstrate that the proposed scheme can achieve the design goals against various attacks.

\subsubsection{\textbf{Mutual Authentication}}

During the access delegation phase, $MD_i$ generates an authorized token $d_i$ for $DT_j$ using DT's public key and $k_{SEAF_i}$ (shared between $MD_i$ and $AMF$). $MD_i$ signs $d_i$ as $u_i$ and encrypts it with $k_{ij}$ before sending to $DT_j$. $DT_j$ then signs $u_i$ and encrypts it using $AMF_1$'s public key. This ensures only the legitimate $AMF_1$ can decrypt $u_i$. If $u_i$ is compromised, $AMF_1$ can detect this by recomputing $t_i=H_5(k_{SEAF_i}\oplus N_i)$ using the received $GUTI_i$ and verifying that $u_i=e(sk_i\cdot d_i, P)=e(sk_i\cdot P, d_i)=e(pk_i,t_i\cdot pk_j)$ fails with an attacker's public key. This allows $AMF_1$ to authenticate $DT_j$. Additionally, $DT_j$ verifies $AMF_1$'s legitimacy and the access delegation $\delta_j$ by computing $\delta_j P=h_1\cdot bpk_{a1}+h_2\cdot R_j+pk_{a1}$.

In the handover authentication phase, $gNB_2$ verifies $DT_j$'s legitimacy using its public key to check signature $\lambda_j$ (equation~\ref{eq:2}). Upon verification, $gNB_2$ accepts that $DT_j$ is authorized by $AMF_1$. $gNB_2$ then generates $MAC_2$ using $sk_{g2}$ and $h_5$ using $TIK_i^*$ (derived from $k_{gNB}^*$). $DT_j$ verifies $MAC_2$ using $gNB_2$'s public key $pk_{g2}$, confirming $gNB_2$'s legitimacy and the correctness of $C_{g2}$ and $h_5$. $DT_j$ then encrypts $h_5$ in $MAC_3$ using $k_{ij}$ and sends it to $MD_i$. After verifying $MAC_3$, $MD_i$ accepts $C_{g2}$ and $h_5$ as legitimate and authenticates $gNB_2$ by validating $h_5$. Finally, $gNB_2$ verifies $MD_i$ by checking $MAC_4$, which is possible because only legitimate $MD_i$ and $gNB_2$ can compute $K_i$ and $k_{gNB}^*$.
\begin{equation}
  \begin{split}
\lambda_jP&=(\delta_j\cdot h_3+b_{j}\cdot h_4)\cdot P\\
&=\delta_j\cdot  h_3\cdot P+h_4\cdot B_{j} \\
&=h_3\cdot(pk_{a1}+ h_1\cdot x_{a1}\cdot P+ h_2\cdot R_j)+h_4\cdot B_j\\
&=h_3\cdot pk_{a1}+h_1 h_3\cdot bpk_{a1}+h_2h_3\cdot R_{j}+h_4\cdot B_j
\label{eq:2}
\end{split}
\end{equation}

\subsubsection{\textbf{Key Negotiation}}
In the proposed scheme,  $MD_i$ transmits $A_i$ to $gNB_2$, and $gNB_2$ delivers $C_{g2}$ to $MD_i$ via $DT_j$. Subsequently, they can individually calculate the secret value $K_i$ and derive the temporary session key $k_{gNB}^*$. It is infeasible for any attackers or even $DT_j$ to compute $K_i$ without knowing the secret values $a_i$, $sk_i$, $c_{g2}$, and $sk_{g2}$, even if $A_i$ and $C_{g2}$ have been intercepted according to ECDLP and ECDHP.  The correctness of $K_i$ is shown in equation \ref{eq:tk}.
\begin{equation}
  \begin{split}
K_i&= a_i\cdot sk_i \cdot C_{g2}\\
&=a_i\cdot sk_i\cdot c_{g2}\cdot sk_{g2}\cdot  P\\
&=c_{g2}\cdot sk_{g2}\cdot  a_i\cdot  sk_i\cdot P\\
&=c_{g2}\cdot  sk_{g2}\cdot a_i \cdot  pk_i\\
&=c_{g2}\cdot  sk_{g2}\cdot  A_{i}
\label{eq:tk}
\end{split}
\end{equation}

\subsubsection{\textbf{Anonymity and Unlinkability}} In the proposed scheme,  true identity of the $MD_i$ is hidden in temporary value $GUTI_i$, which is used by $DT_j$ to initiate the handover authentication with gNBs. Only the 5GC can reveal the true identity $SUPI_i$ of $MD_i$. In addition, $MD_i$ generates different  $TID_i$ with gNBs in intra-domain handover and updates the $GUTI_i$ with  AMFs in inter-domain handover. It is difficult for an adversary to determine that the two temporary identities belong to the same $MD_i$ when eavesdropping on the wireless channel. 

\begin{table*}[htb]
\centering
\renewcommand{\arraystretch}{1.15}
\begin{threeparttable}
\caption{Functionality Comparison}
\begin{tabular}{c|c c c c c c c c c}
\hline
\diagbox{Scheme}{Functionality}  &MAKA & Anonymity & Unlinkability & Traceability & Key Confirmation & PFS & PBS & KEF & ELS \\
\hline
\hline
5G-AKA  \cite{3gpp.36.331} &$\checkmark$ &$\checkmark$ &$\times$      &$\checkmark$  &$\times$             &$\times$       &$\times$      &$\times$    &$\times$\\
Lai et al.'s \cite{lai2022novel}     &$\checkmark$  &$\checkmark$  &$\times$     &$\checkmark$     &$\times$      &$\checkmark$   &$\checkmark$  &$\times$     &$\times$     \\
Ma et al.'s  (I) \cite{ma2019ftgpha}  &$\checkmark$&$\checkmark$  &$\times$     &$\checkmark$     &$\times$     &$\times$   &$\times$  &$\times$     &$\times$     \\
Ma et al.'s (II) \cite{ma2019ftgpha}  &$\checkmark$&$\checkmark$  &$\times$     &$\checkmark$     &$\times$     &$\checkmark$   &$\checkmark$  &$\times$     &$\times$     \\
Cao et al.'s \cite{cao2019cppha}  &$\checkmark$&$\checkmark$    &$\checkmark$      &$\checkmark$  &$\checkmark$             &$\times$       &$\times$      &$\times$    &$\times$\\
Zhang et al.'s \cite{zhang2019robust}    &$\checkmark$&$\checkmark$  &$\times$     &$\checkmark$  &$\checkmark$              &$\checkmark$  &$\checkmark$      &$\checkmark$ &$\checkmark$\\
Yan et al.'s \cite{yan2022efficient}   &$\checkmark$ &$\checkmark$   &$\checkmark$    &$\checkmark$  &$\times$ &$\checkmark$ &$\checkmark$ &$\checkmark$ &$\times$\\ \hline
Gupta et al.'s \cite{gupta2018proxy}  &$\checkmark$ &$\checkmark$ &$\times$   &$\times$  &$\checkmark$  &$\checkmark$ &$\checkmark$ &$\checkmark$ &$\times$\\
He et al.'s \cite{2017r17}   &$\checkmark$&$\checkmark$ &$\times$  &$\checkmark$  &$\times$ &$\checkmark$ &$\times$ &$\times$ &$\times$ \\
Wang et al.'s \cite{2021blockchain} &$\checkmark$  &$\checkmark$ &$\times$   &$\checkmark$  &$\times$  &$\checkmark$ &$\checkmark$ &$\checkmark$ &$\times$\\
Li et al.'s \cite{li2021seccdv}  &$\checkmark$  &$\checkmark$ &$\times$   &$\checkmark$  &$\checkmark$  &$\checkmark$ &$\checkmark$ &$\checkmark$ &$\times$\\\hline
Ours  &$\checkmark$ &$\checkmark$ &$\checkmark$ &$\checkmark$   &$\checkmark$  &$\checkmark$  &$\checkmark$ &$\checkmark$ &$\checkmark$\\
\hline

\end{tabular}

\begin{tablenotes}
\item 1. MAKA: MD and base station achieve mutual authentication and key agreement; 2. Anonymity: MD's true identity should be protected from base stations and adversaries; 3. Unlinkability: The adversary cannot distinguish that more different messages originate from the same MD. 4. Traceability: The true identity of MD should be revealed when misbehavior happens;   5. Key Confirmation: MD and base station should confirm that each other has generated the secret key successfully; 6. PFS: Perfect Forward Secrecy; 7. PBS: Perfect Backward Secrecy; 8: KEF: Key Errow Freshness; 9: ESL: Ephemeral Secret Leakage.
\item 2. $\checkmark$ represents the functionality is achieved; $\times$ represents the functionality is not achieved.
\end{tablenotes}
\end{threeparttable}
\label{tb:1}
\end{table*}

\subsubsection{\textbf{Traceability}}
There are malicious $MD_i$ that can apply and deploy the corresponding compromised digital twin, using intercepted information such as $u_i$, to engage in unauthorized access delegation with AMFs or initiate handover processes with gNBs. During the verification phase, if any suspicious behavior is detected,   traceability measures can be activated through $UDM$ and $AUSF$. Using the $ID_j$ identifier, $UDM$ can search the corresponding $U_j$, enabling $AUSF$ to compute the $SUPI_i=ID_j\oplus H_1(sU_j)$ to reveal the true identity of $MD_i$. $AUSF$ and $UDM$ can further revoke the public key of $DT_j$ and imposing the application of $MD_i$ in the network.
\subsubsection{\textbf{Perfect Forward/Backward Secrecy (PFS/PBS)}} In the proposed scheme, the secret value $K_i=a_i\cdot sk_i\cdot C_{g2}$ or $K_i=c_{g2}\cdot sk_{g2}\cdot A_{i}$ and temporary session key $k_{gNB}^*=H_2(K_i,GUTI_i,ID_{g2})$. Since $a_i$ and $c_{g2}$ are chosen randomly in each session, even if the attacker can obtain the long-term keys $sk_i$ and $sk_{g2}$, it is difficult to derive the previous secret value $K_i$ and session key $k_{gNB}^*$, thus achieving PFS. Moreover, since future sessions will use new random values $a_i$ and $c_{g2}$ that are independent of previous sessions, even if an attacker obtains previous session keys $k_{gNB}^*$, they cannot derive or compromise the keys for subsequent sessions, thus achieving PBS.
\subsubsection{\textbf{Key Escrow Freeness (KEF)}} In the system initialization phase, $MD_i$, $DT_j$, gNB independently generate their private key. In addition, although AUSF generates partial secret key $x_a$ for AMF, which generates its own private key $sk_a$ and introduces random number $r_j$ in each access delegation so that AUSF cannot derive $\delta_j$ with $x_a$ alone. Therefore, the proposed scheme is a key escrow-free handover authentication protocol.
\subsubsection{\textbf{Ephemeral Secret Leakage (ESL)   Resistance}}  In the proposed scheme, the session key $k_{gNB}^*=H_2(a_i\cdot sk_i \cdot c_{g2}\cdot sk_{g2}\cdot P, GUTI_i, ID_{g2})$  is derived from combination of ephemeral secrets ($a_i$, $c_{g2}$) and long-term keys ($sk_i$, $sk_{g2}$). If an adversary attempts to compute the $k_{gNB}^*$, it requires that both ephemeral and long-term secrets must be leaked simultaneously, which is highly unlikely in practice.  Therefore, our proposed scheme is resistant to ephemeral secret leakage.

\subsubsection{\textbf{Protocol Attack Resistance}} In the proposed scheme, the confidential data including  $u_i$ is encrypted by the $k_{ij}$ between $MD_i$ and $DT_j$ and the  access delegation $\delta_j$ and $u_i$ is encrypted by  the each other's public key between $DT_j$ and $AMF_1$ to resist \textbf{\textit{eavesdropping}}. The random session number $N_i$ generated by $MD_i$ and the timestamp $TS$ added in each session is to defend the replay attack from the malicious attacker who intercepts the previous data.  The hash function, message authentication code, and digital signature are employed in each session to resist \textbf{\textit{impersonation attack}} and \textbf{\textit{man-in-the-middle attack}}. In addition, our scheme supports batch verification which can alleviate the  \textbf{\textit{DoS attacks}} for gNBs.

\section{Performance Evaluation}
In this section, we analyze the security functionality and performance including signaling, communication, and computation overheads of the proposed scheme between MD and gNB in the intra-domain handover authentication phase, and compare it with other related handover authentication schemes.

\subsection{Security Functionality}
Table II presents the comparison of the security functionalities between our proposed scheme and other relevant handover authentication schemes. 5G-AKA \cite{3gpp.36.331} is standard scheme as defined by 3GPP, Lai et al.'s scheme \cite{lai2022novel}, Ma et al.'s scheme I and scheme II \cite{ma2019ftgpha}, Cao et al.'s scheme\cite{cao2019cppha}, Zhang et al.'s scheme \cite{zhang2019robust} and  Yan et al.'s scheme \cite{yan2022efficient} are handover scheme in 5G related scenarios; Gupta et al.'s scheme \cite{gupta2018proxy}, He et al.'s scheme \cite{2017r17},  Wang et al.'s scheme \cite{2021blockchain} and Li et al.'s scheme \cite{li2021seccdv} are handover scheme in other wireless network scenarios. Through the analysis, all schemes can achieve the basic goal of mutual authentication and key agreement between MD and base station. Regarding identity privacy preservation, all schemes support the anonymity of the MD and the traceability of the central server or core network, except for schemes \cite{gupta2018proxy}. However, only Ma et al.'s scheme II \cite{ma2019ftgpha}, Cao et al.'s scheme \cite{cao2019cppha}, Yan et al.'s scheme \cite{yan2022efficient} and ours support unlinkability to prevent tracking by adversaries on the wireless channel.  Cao et al.'s scheme \cite{cao2019cppha}, Zhang et al.'s scheme \cite{zhang2019robust}, Gupta et al.'s scheme \cite{gupta2018proxy}, Li et al.'s scheme \cite{li2021seccdv}, and ours support key confirmation.  In the context of secret key security, 5G-AKA \cite{3gpp.36.331}, Ma et al.'s scheme I \cite{ma2019ftgpha} and Cao et al.'s scheme\cite{cao2019cppha} fail to achieve PFS, PBS and KEF. He et al.'s scheme \cite{2017r17} can not achieve PBS and KEF. Ma et al.'s scheme II \cite{ma2019ftgpha} and  Lai et al.'s scheme \cite{lai2022novel} can not achieve KEF.  Additionally, only Zhang et al.'s scheme \cite{zhang2019robust} and ours can resist ESL attack.

\subsection{Signaling Overhead}
\begin{figure}[t]
  \centering
  \includegraphics[width=0.3 \textwidth]{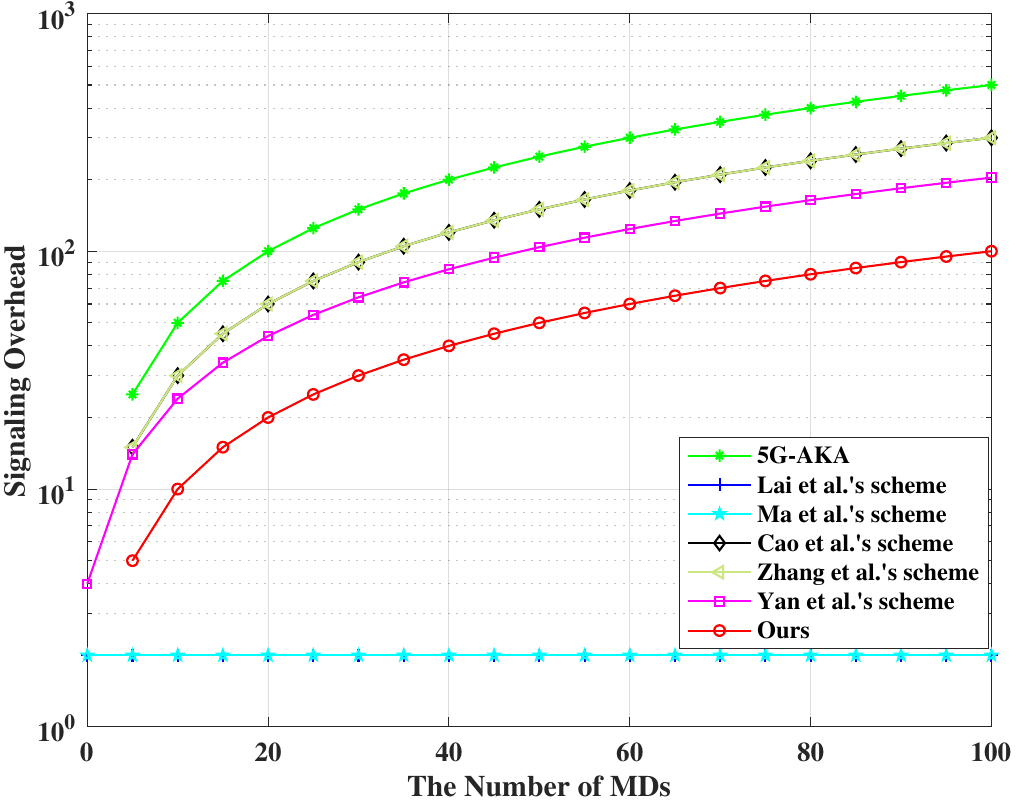}
  \caption{Signaling Overhead in Handover Authentication}
  \label{fig:trans}
\end{figure}

Signaling overhead refers to the costs of transmitting signaling messages to achieve handover authentication between the MD and the base station.  We use $a$ to represent the overhead of transmitting a signaling message between the MD and the base station and $n$ to represent the number of MDs. As shown in Table \ref{tab:2}, we compare the signaling overheads with other 5G-related schemes and illustrate this comparison in Fig.~\ref{fig:trans}. By comparison, our scheme demonstrates lower signaling overhead compared to most existing schemes, except for Ma et al.'s scheme \cite{ma2019ftgpha} and Lai et al.'s scheme \cite{lai2022novel}, which utilize a group leader to generate aggregated authentication codes for base stations on behalf of other MDs. In our proposed scheme, digital twin have replaced MD to transmit some necessary messages to the base station through dedicated API wired channels. This not only saves wireless resources but also reduces the signaling overhead. The MD only needs to send one message to the base station to complete authentication. In fact, this message serves as the key confirmation to complete the entire handover authentication procedure between the MD and the target base station. 
\begin{table}[htb]
\centering
\renewcommand{\arraystretch}{1.15}
  \caption{Comparison of Signaling Overhead}
\begin{tabularx}{6.cm}{c|c}
\hline
\textbf{Schemes}       & \textbf{Signaling Overhead} \\
\hline
\hline
5G-AKA \cite{3gpp.36.331} & $5an$                              \\
Lai et al.'s \cite{lai2022novel}     &$2a$\\
Ma et al.'s \cite{ma2019ftgpha}      &$2a$\\
Cao et al.'s \cite{cao2019cppha}     &$3an$\\
Zhang et al.'s \cite{zhang2019robust}   &$3an$\\
Yan et al.'s \cite{yan2022efficient}   &$2an+4a$ \\\hline
Gupta et al.'s \cite{gupta2018proxy} &$3an$\\
He et al.'s \cite{2017r17}            &$3an$\\
Wang et al.'s \cite{2021blockchain}   &$3an$\\
Li et al.'s \cite{li2021seccdv}     &$3an$\\
\hline
Ours                    &$an$ \\
\hline
\end{tabularx}
\label{tab:2}
\end{table}


\begin{table}[htb]
 \captionsetup{justification=centering}
\centering
\renewcommand{\arraystretch}{1.15}
\caption{Computation Costs of the Primitive Cryptography Operations (ms)}
\begin{tabular}{c|c c c c c c }
\hline
          & $T_p$ & $T_e$   &$T_m$   &$T_r$   &$T_h$        \\
\hline
MD            &2.87   &0.225    &0.203  &0.127  &0.0013  \\
Base Station  &0.762  &0.034    &0.03    &0.019  &0.0008   \\
\hline
\end{tabular}
\label{ta:3}
\end{table}
\subsection{Computation Overhead}

We present the main computation costs of cryptographic operations for both the MD and base station in Table \ref{ta:3}. Testing was performed on an Intel i5-2500 @ 3.30 GHz (MD) and Intel i7-6600U @ 2.60 GHz (base station, including gNB, eNB, and other access points) \cite{2021blockchain}. The operations are denoted as $T_p$ for pairing operation, $T_e$ for modular exponentiation, $T_m$ for elliptic curve scalar multiplication, $T_r$ for RSA signature verification, and $T_h$ for hash operation. Lightweight operations (symmetric encryption, XOR, point addition) are omitted due to negligible costs. We analyze computation overhead in two scenarios: the normal scenario considers all computations during handover authentication, and the optimized scenario only considers computations after MD enters the target base station's coverage area.

\begin{figure}[tbp]
   \centering
   \subfloat[Normal Scenario]{
       \includegraphics[width=0.226 \textwidth]{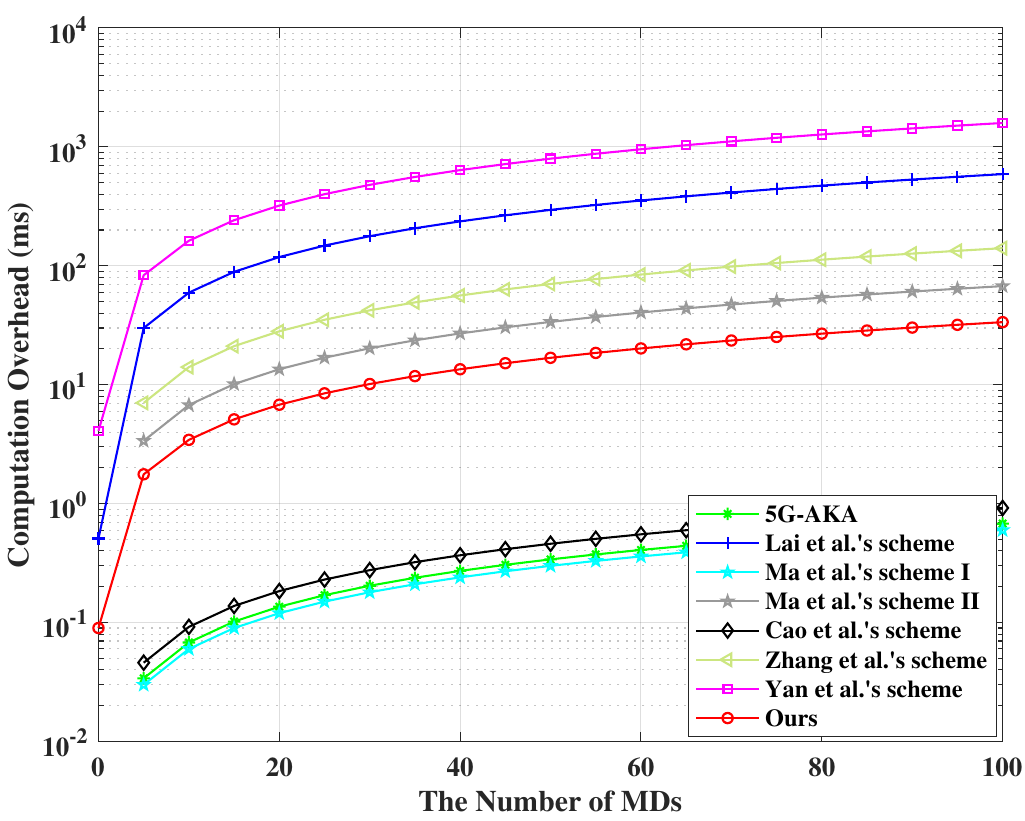}
    }
    \hfill
    \subfloat[Optimized Scenario]{
        \includegraphics[width=0.226 \textwidth]{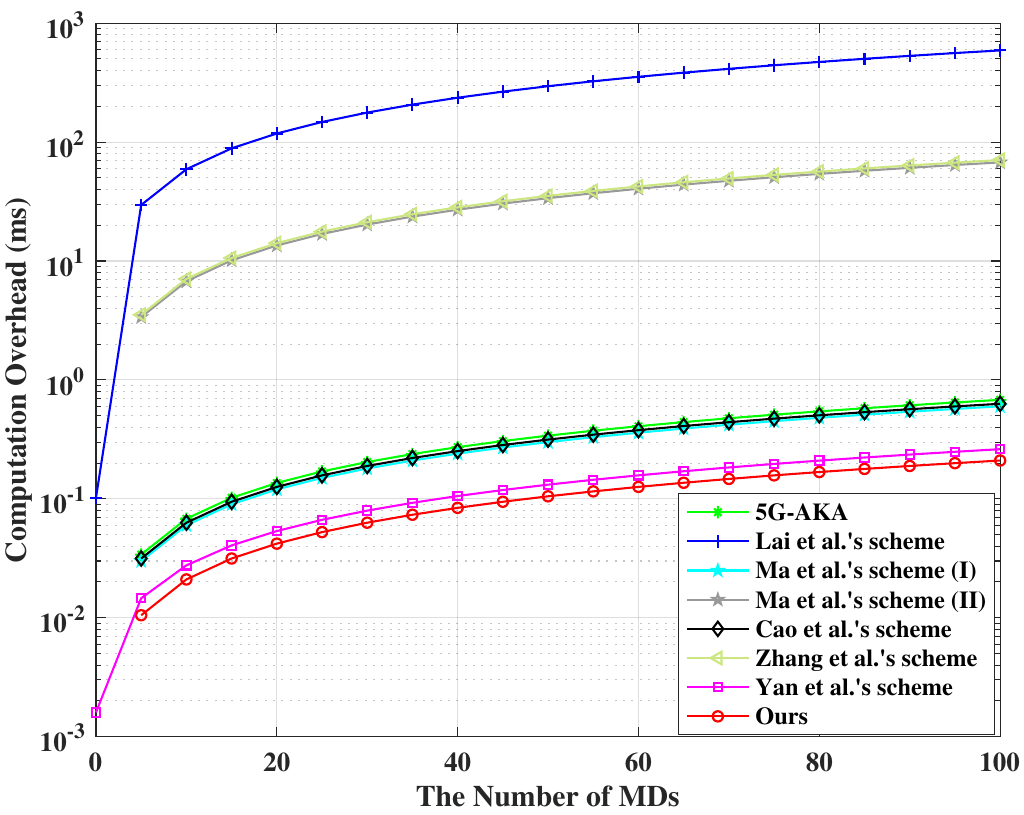}
    }
 
    \caption{Computation Overhead in Handover Authentication}
      \label{fig:comp}
\end{figure}

In Table \ref{ta:4}, we compare our scheme with other handover authentication schemes and visualize the total computation overhead comparison in Fig.~\ref{fig:comp}. Through comprehensive analysis, our results reveal that in normal scenarios, our scheme exhibits slightly higher computation overhead compared to 5G-AKA \cite{3gpp.36.331}, Ma et al.'s scheme (I) \cite{ma2019ftgpha}, and Cao et al.'s scheme \cite{cao2019cppha}. This increased overhead stems from two primary factors: these existing schemes utilize lightweight symmetric encryption for handover authentication, while our scheme requires the target base station to perform additional multiplication operations to verify the digital twin's legitimacy. However, in optimized scenarios, our scheme demonstrates significantly lower computation overhead than all other compared schemes. The MD and target base station each only need to perform a single hash operation to complete the entire handover authentication process. This remarkable efficiency is achieved because the digital twin proactively completes most computational tasks while the MD remains within the source base station's coverage, pre-establishing authentication and key agreement with the target base station.

\begin{table*}[htb]
\centering
\renewcommand{\arraystretch}{1.25}
\caption{Comparison of Computation Overhead in  Handover Authentication (ms)}
\resizebox{\textwidth}{!}{
\begin{tabular}{|c|c|c|c|c|c|c|}
\hline
\multicolumn{1}{|c|}{Schemes} & \multicolumn{1}{c|}{$T_{MD}$} & \multicolumn{1}{c|}{$T_{BS}$} & \multicolumn{1}{c|}{$T_{tot}$} & \multicolumn{1}{c|}{$T_{MD-opt}$} & \multicolumn{1}{c|}{$T_{BS-opt}$} & \multicolumn{1}{c|}{$T_{tot-opt}$} \\ \hline
\hline
5G-AKA\cite{3gpp.36.331}     & $4nT_h$                           & $2nT_h$                &$0.016n$                  & $4nT_h$           & $2nT_h$             &$0.016n$                 \\ \hline
Lai et al.'s \cite{lai2022novel}          & $2T_m+T_h+nT_r+2nT_p$             & $3T_e+nT_m$                  & $5.897n+0.509$           & $nT_r+2nT_p$           & $nT_m+3T_e$               & $5.897n+0.102$            \\ \hline
Ma et al.'s (I)\cite{ma2019ftgpha}   &$5nT_h$                            &$nT_h$                       &$0.007n$                   &$5nT_h$                            &$nT_h$                       &$0.007n$    \\ \hline  
Ma et al.'s (II)\cite{ma2019ftgpha}  &$3nT_m+4nT_h$                      &$2nT_m+nT_h$                 &$0.675n$                    &$3nT_m+4nT_h$                      &$2nT_m+nT_h$                 &$0.675n$   \\ \hline
Cao et al.'s \cite{cao2019cppha}          &$4nT_h$                             &$5nT_h$                  &$0.0092n$                     &$3nT_h$                 &$3nT_h$                    &$0.0063n$                   \\ \hline
Zhang et al.'s \cite{zhang2019robust}       & $6nT_m+4nT_h$                          & $6nT_m+5nT_h$                & $2.001n$                 & $3nT_m+2nT_h$          & $3nT_m+6nT_h$             & $1.067n$        \\ \hline
Yan et al.'s \cite{yan2022efficient}      & $5nT_p+7nT_m+8nT_h$                    & $(2n+9)T_m+4(n+1)T_h+5T_p$   & $15.781n+0.063$          & $2nT_h$                & $2nT_h$                   & $0.004n$           \\ \hline\hline
Gupta et al.'s \cite{gupta2018proxy}        & $7nT_m+7nT_h$                          & $12nT_m+7nT_h$               & $1.878n$                 & $3nT_m+4nT_h$          & $3nT_m+4nT_h$             & $0.754n$         \\ \hline
He et al.'s \cite{2017r17}               & $3nT_e+4nT_m$                          & $3nT_p+nT_e+2T_m$            & $3.64n$                  & $nT_e$                 & $3nT_p+nT_m$              & $2.541n$           \\ \hline
Wang et al.'s \cite{2021blockchain}        & $2nT_m+2nT_h$                          & $7nT_m+5nT_h$                & $0.623n$                 & $2nT_m+2nT_h$          & $nT_h$                    & $ 0.409n $       \\ \hline
Li et al.'s \cite{li2021seccdv}          & $14nT_m+5nT_h$                         & $8nT_m+5nT_h$                & $3.093n$                 & $4nT_m+3nT_h$          & $7nT_m+4nT_h$             & $1.029n$                        \\ \hline\hline
Ours                         & $nT_m+7nT_h$                      & $(5n+3)T_m+11nT_h$            & $0.372n+0.09$                & $nT_h$                 & $nT_h$                    & $0.002n$                          \\ \hline
\end{tabular}
}
\label{ta:4}
\end{table*}

\subsection{Communication Overhead}
We further analyze the communication overhead, which includes the size of the necessary parameters forwarded between the MD and the target base station during the handover authentication process. In order to achieve the same security level of key strength, it is assumed that the encryption and decryption key length of AES is 128 bits, according to the  National Institute of Standards and Technology (NIST) standard. In addition, we assume the sizes of the elements in the cycle group $G$ is 256 bits, and the key size is 3072 bits for both RSA (integer-factorization cryptography) and finite-field cryptography-based public keys. The output size is 128 bits for hash values, message authentication codes, random numbers, and identities. The MD's capability and proxy warrant are 80 bits each, the timestamp is 32 bits and the sequence number is 48 bits. Considering that the Lai et al.'s scheme \cite{lai2022novel}, Ma et al.'s scheme\cite{ma2019ftgpha}, and Yan et al.'s scheme \cite{yan2022efficient} are group-based handover authentication, we have disregarded the communication overhead within the group. 

\begin{figure}[t]
  \centering
  \includegraphics[width=0.3 \textwidth]{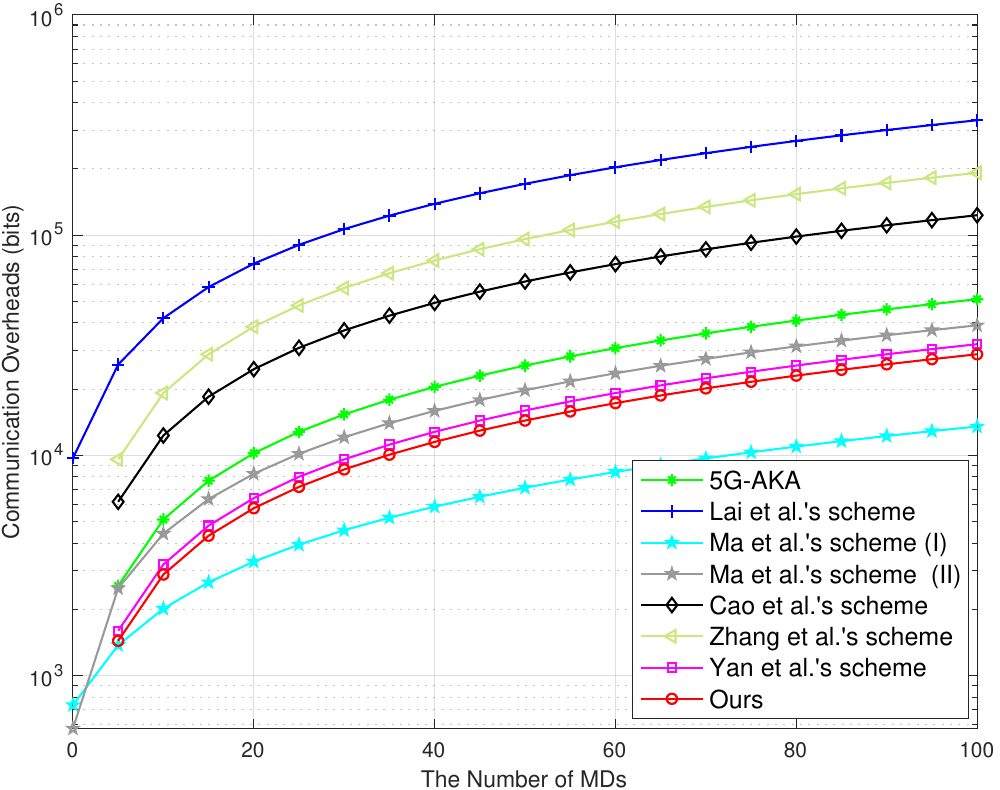}
  \caption{Communication Overhead in Handover Authentication}
  \label{fig:comm}
\end{figure}

As shown in Table \ref{table:2}, for the standard 5G-AKA \cite{3gpp.36.331}, the total communication overhead is $512n$ bits, where $n$ represents the number of MDs. This overhead comprises a 128-bit random number $RAND$, 128-bit authentication token $AUTN$, 128-bit concealed subscriber identity $SUCI$, and 128-bit response value $RES^*$.  In Lai et al.'s scheme \cite{lai2022novel}, the total communication overhead is $3328n+9728$ bits which is based on finite-field cryptography. In Ma et al.'s  (I) \cite{ma2019ftgpha}, the total communication overhead is $128n+768$  and in Ma et al.'s  (II) \cite{ma2019ftgpha} is $384n+976$ where the latter uses the ECC encryption. In Cao et al.'s \cite{cao2019cppha}, the total communication overhead is $1184n$ including the necessary capability generated by AHM. In Zhang et al.'s \cite{zhang2019robust}, the total communication overhead is $1856n$ bits. In Yan et al.'s \cite{yan2022efficient}, the total communication overhead is $320n+1024$ bits including the pre-handover messages, and the handover authentication executed between each vehicle and the target base station. In our scheme, on the one hand, the essential parameters from the MD to the target base station for authentication are transmitted by digital twin using wired communication. On the other hand, the DT transmits the essential 544 bits pre-handover parameters from source base station to the MD for authentication via the data plane. This approach conserves wireless resources at the control plane, requiring the MD to send only a 288-bit key confirmation to complete the handover authentication process with the target base station.  As shown in Fig. \ref{fig:comm},  it can be observed that our scheme has higher communication overhead compared to Ma et al.'s (I) \cite{ma2019ftgpha} but is better than other schemes, which helps in saving more wireless resources.

\begin{table}[htb]
\captionsetup{justification=centering}
\centering
\renewcommand{\arraystretch}{1.15}
  \caption{Communication Overhead in Handover Authentication (bits)}
\begin{tabular}{c|ccc}
\hline
\textbf{Schemes}       & \textbf{Uplink}      &\textbf{Downlink}       &\textbf{Total}\\ \hline
\hline
5G-AKA\cite{3gpp.36.331}     & $256n$                           & $256n$                &$512n$                                 \\ 
Lai et al.'s \cite{lai2022novel}         & $3328n+3328$              & $6400$                   & $3328n+9728$                       \\ 
Ma et al.'s  (I)\cite{ma2019ftgpha}   &$128n+384$                            &$384$                       &$128n+768$                     \\ 
Ma et al.'s  (II)\cite{ma2019ftgpha}  &$384n+464$                      &$512$                 &$384n+976$                      \\ 
Cao et al.'s \cite{cao2019cppha}          &$640n$                             &$424n$                  &$1184n$                                      \\ 
Zhang et al.'s \cite{zhang2019robust}       & $928n$                          & $928n$                & $1856n$                   \\ 
Yan et al.'s \cite{yan2022efficient}      & $288n+512$                    & $32n+512$              &$320n+1024$    \\ \hline
Gupta et al.'s \cite{gupta2018proxy}      &$1104n$                     & $1104n$              &$2208n$    \\
He et al.'s \cite{2017r17}                &$1128n$                     & $384n$              &$1512n$    \\
Wang et al.'s \cite{2021blockchain}    & $672n$                       & $832n$               &$1504n$   \\
Li et al.'s \cite{li2021seccdv}       & $1712n$                       &$1008n$              &$2720n$ \\\hline
Ours      & $288n$                    & $-$   & $288n$                  \\ \hline

\end{tabular}
\label{table:2}
\end{table}


\subsection{Performance with Unknown Attacks}

Although our scheme can resist common attacks discussed in Section V, unknown attacks may still occur unpredictably \cite{ma2019ftgpha}. In our scheme, these unknown attacks could disrupt DT operation and data transmission to the MD, preventing advance handover authentication and key negotiation between the MD and base station. We analyze our protocol's performance against unknown attacks in terms of signaling, computational, and communication overhead \cite{cao2019cppha}. This includes analyzing DT's computation overhead for parameter processing and communication overhead for data transmission to MD through the data plane. Focusing on communication overhead analysis, we express the average communication overhead $Com_{avg}$ under attacks in equation \ref{eq:com}, where $Com_{fail}$ represents unsuccessful authentication overhead under unknown attacks and $Com_{succ}$ represents successful authentication overhead under known attacks \cite{ma2019ftgpha}. We use $p_{fail}$ to denote the probability of an unknown attack occurring and $p_{succ}=1-p_{fail}$ for success probability. Additionally, $Com_{fail}=\sum_{i=1}^N Com_{i}\times q$ where $N$ is the total number of authentication messages, $q = 1/N$ is the probability of an unknown attack at step $i$, and $Com_{i}$ is the total communication overhead before an attack occurs at step $i$.
\begin{equation}
\begin{split}
Com_{avg}=\frac{Com_{fail}\times p_{fail}+Com_{succ}\times p_{succ}}{p_{succ}}
\label{eq:com}
\end{split}
\end{equation}

\begin{figure*}[htbp]
    \centering
    \subfloat[Signaling Overhead]{
        \includegraphics[width=0.29\textwidth]{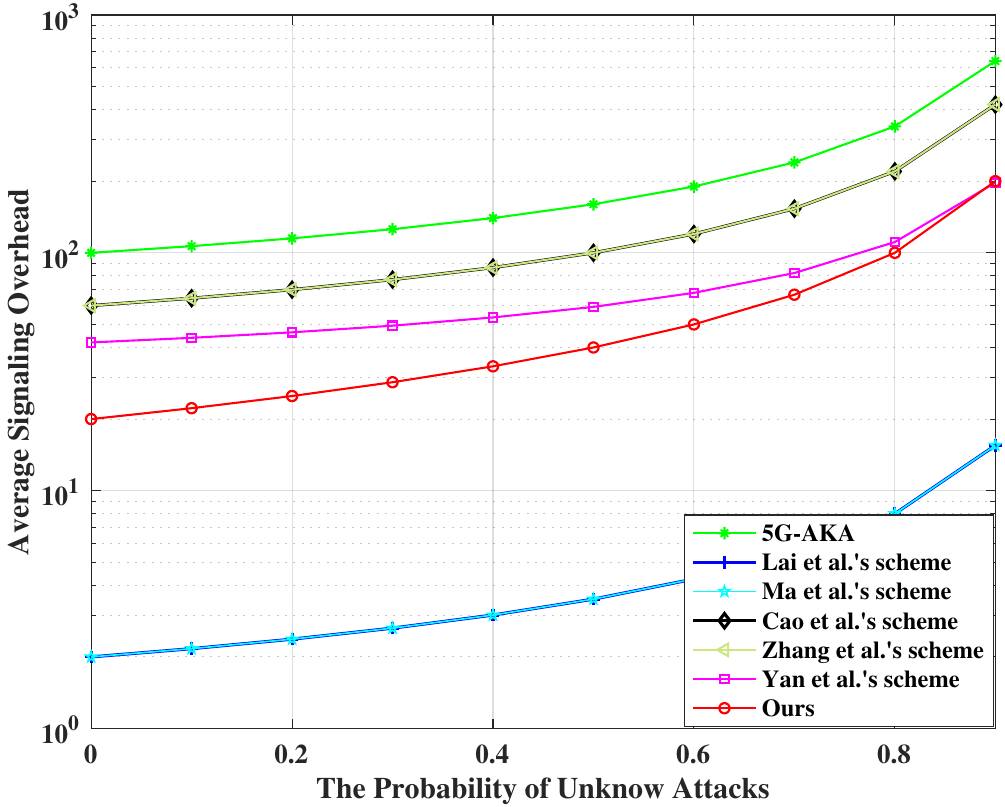}
    }
    \subfloat[Computation Overhead in Normal Scenario]{
        \includegraphics[width=0.29\textwidth]{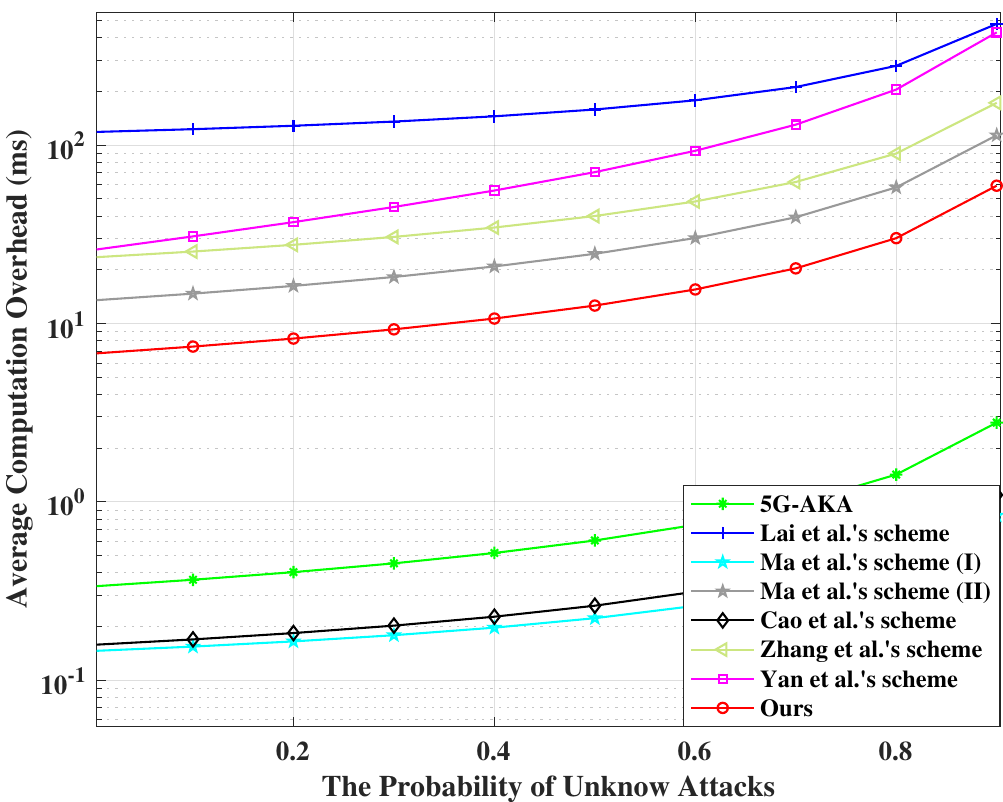}
    }
    \\
    \subfloat[Computation Overhead in Optimized Scenario]{
        \includegraphics[width=0.29\textwidth]{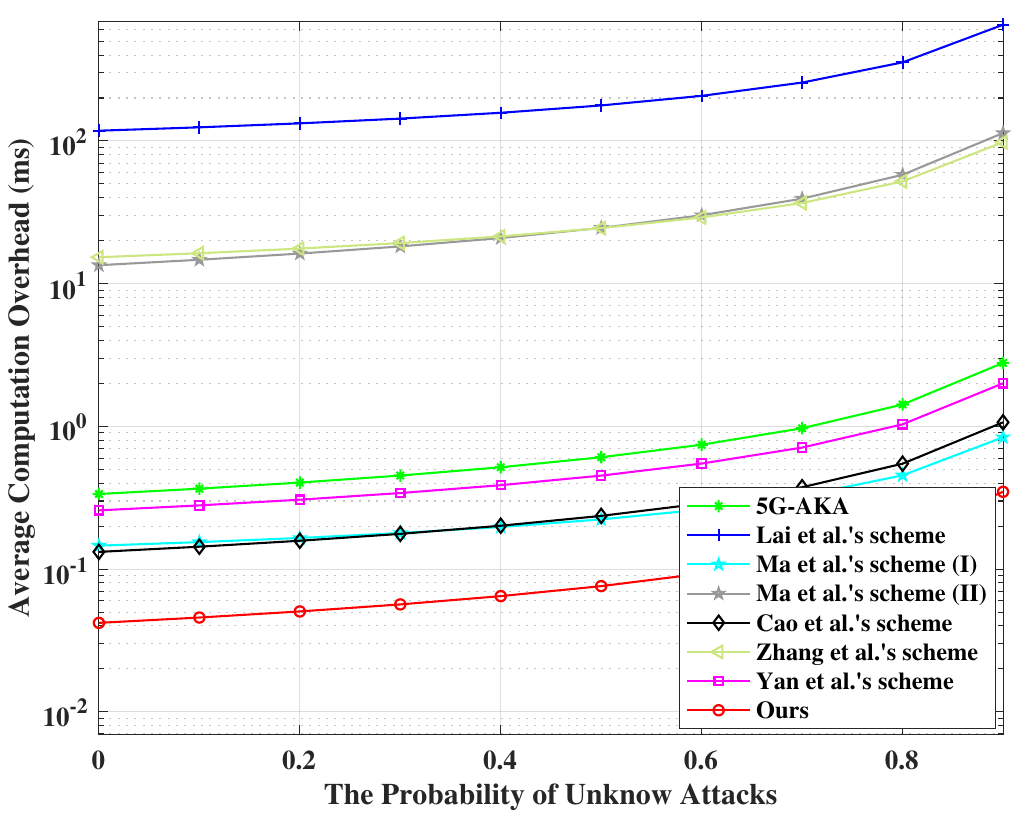}
    }
    \subfloat[Communication Overhead]{
        \includegraphics[width=0.29\textwidth]{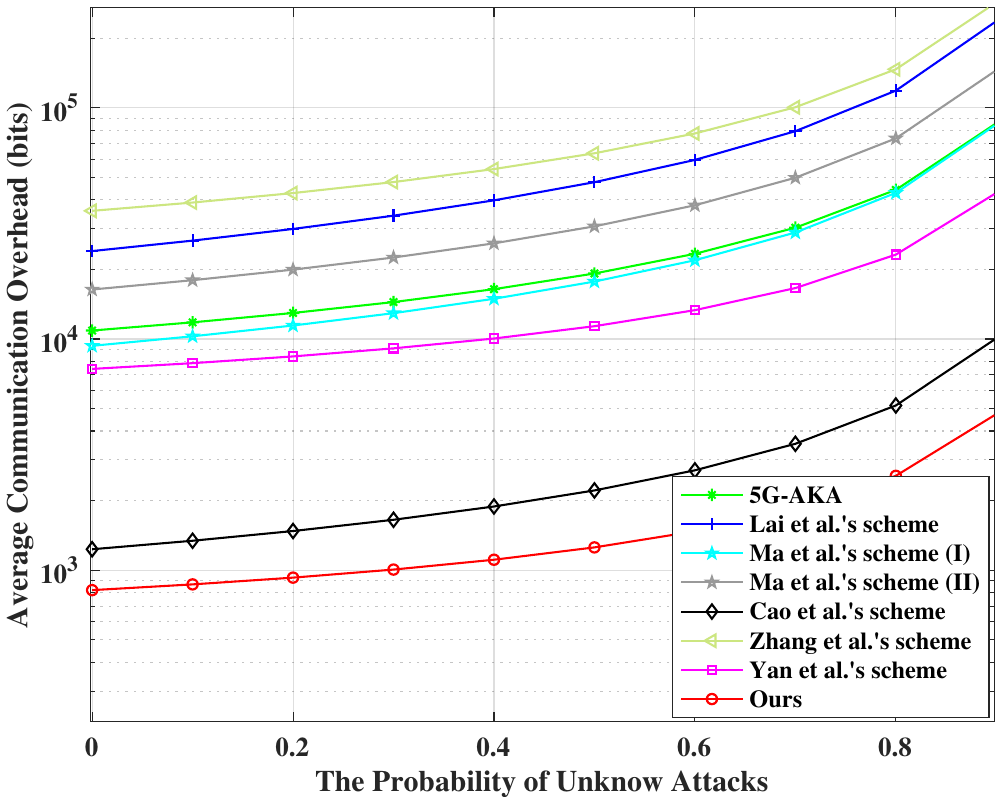}
    }
    \caption{ Comparison of the Performance under Unknown Attacks.}
    \label{fig:unknown}
\end{figure*}

As shown in Fig. \ref{fig:unknown}, we compare average signaling, computation, and communication overhead with other 5G handover authentication schemes under unknown attacks. For group-based schemes \cite{ma2019ftgpha, lai2022novel, yan2022efficient}, we assume 20 MDs per group and consider unknown attacks affecting wireless communication between group members. We also assume DT has equivalent computation capability to the base station (Table III).
In Fig. \ref{fig:unknown}(a), our scheme's signaling overhead under unknown attacks is higher than \cite{ma2019ftgpha} and \cite{lai2022novel} but lower than other schemes. As attack probability increases, our scheme's signaling overhead converges with \cite{yan2022efficient}.
Fig. \ref{fig:unknown}(b) shows our scheme's average computation overhead under normal scenarios exceeds \cite{ma2019ftgpha}, \cite{cao2019cppha}, and \cite{3gpp.36.331} but remains lower than other schemes, primarily due to DT's role in generating and verifying handover requests.
Fig. \ref{fig:unknown}(c) demonstrates our scheme achieves better average computation overhead in optimal scenarios compared to other schemes, benefiting from DT's advanced processing of complex calculations.
Fig. \ref{fig:unknown}(d) reveals that our scheme maintains superior average communication overhead even when accounting for encrypted data and message authentication code transmission from DT to MD via the data plane.


\subsection{Comprehensive Discussion}
Our scheme shows slightly higher signaling overheads compared to Lai et al.'s scheme \cite{lai2022novel} and Ma et al.'s scheme \cite{ma2019ftgpha}, and higher computation overhead compared to standard 5G-AKA \cite{3gpp.36.331}, Ma et al.'s scheme (I) \cite{ma2019ftgpha}, and Cao et al.'s scheme \cite{cao2019cppha}. However, it achieves better communication overhead, conserving wireless channel resources. While Ma et al.'s scheme \cite{ma2019ftgpha} is limited to predictable railway scenarios, our scheme adapts to complex urban environments, enabling proactive session key negotiation with future base stations through real-time trajectory analysis, suitable for various mobile scenarios and reducing authentication computational burden regardless of base station distribution patterns. Moreover, our scheme provides enhanced security functionality while maintaining lower overheads between MD and gNB.

Although DT is crucial in our scheme, it faces various potential attacks in real-world scenarios that could impact handover authentication efficiency. While DT can be deployed in TEE on operator-provided dedicated servers \cite{li2023breaking}, it remains vulnerable to DoS and side-channel attacks. Despite authentication parameters being encrypted by $k_{ij}$, wireless communication faces potential disruption attacks between MD and DT. Even secure interfaces between DT and the base station/core network aren't immune to unauthorized access attempts. Beyond analyzing authentication overhead under unknown attacks, we recommend: periodic $k_{ij}$ updates with continuous authentication between MD and DT, implementing DoS defense and enhanced DT environment security, and regular security audits with improved defensive measures. While these aspects extend beyond our paper's scope, such additional security measures are essential for achieving secure and efficient DT-assisted handover authentication in 5G and beyond.

\section{Conclusion}

In this paper, we propose a novel handover authentication scheme that both supports 5G Intra-AMF and Inter-AMF scenarios with the assistance of digital twin. By our proposed scheme, the digital twin first obtains the delegation from AMF, then it replaces the MD to initiate handover authentication request to the target base station by analyzing the real-time path so that the mobile device and the MD can accomplish mutual authentication and key agreement before attaching. The proposed scheme is analyzed by BAN logic, ProVerif and informal analysis to prove it can achieve several security functionalities. Additionally, our scheme demonstrates better signaling, computation, or communication overhead, attributed to the incorporation of digital twin, in comparison to the majority of existing related schemes. Despite our scheme being slightly higher than a few schemes in performance, we possess significant in terms of security functionality, features support, and application scenarios. In future work, we will analyze and explore employing the digital twin to support massive mobile device access authentication in 5G and beyond.

\bibliographystyle{IEEEtran}
\bibliography{IEEEabrv,Reference}

\begin{IEEEbiography}[{\includegraphics[width=1in,height=1.25in,clip,keepaspectratio]{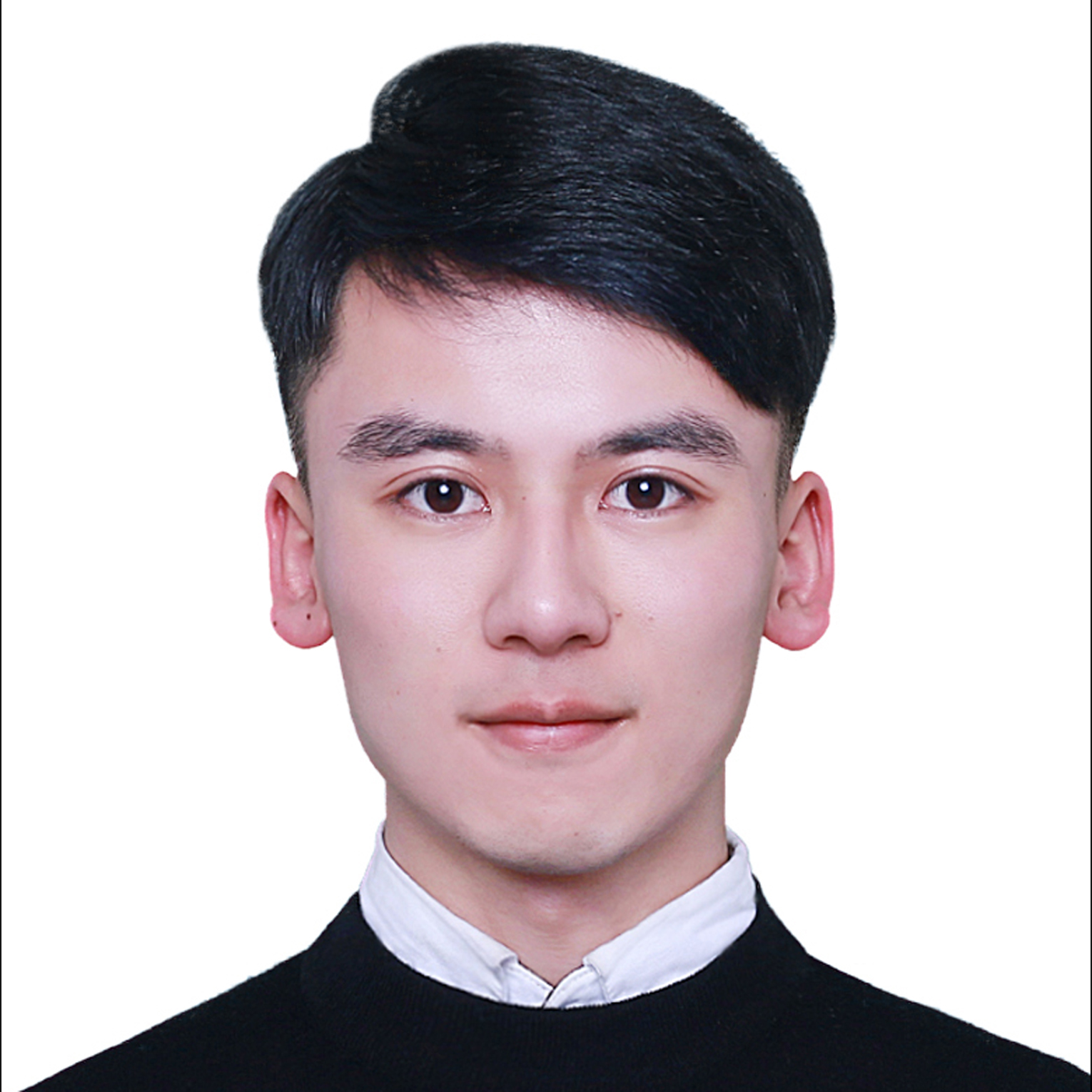}}]{Guanjie Li}
received B.S. degree in communication engineering and  M.S. degree in electronics and communication engineering from the Xi’an University of Posts and Telecommunications, Xi’an, China, in 2016 and 2021, respectively. He is currently a Ph.D. candidate  with the School of Cyber Engineering, Xidian University, Xi'an, China, and a visiting graduate student at the Information Systems Technology and Design Pillar, Singapore University of Technology and Design, Singapore. His research interests include the vehicular network and digital twin.
\end{IEEEbiography}

\begin{IEEEbiography}[{\includegraphics[width=1in,height=1.25in,clip,keepaspectratio]{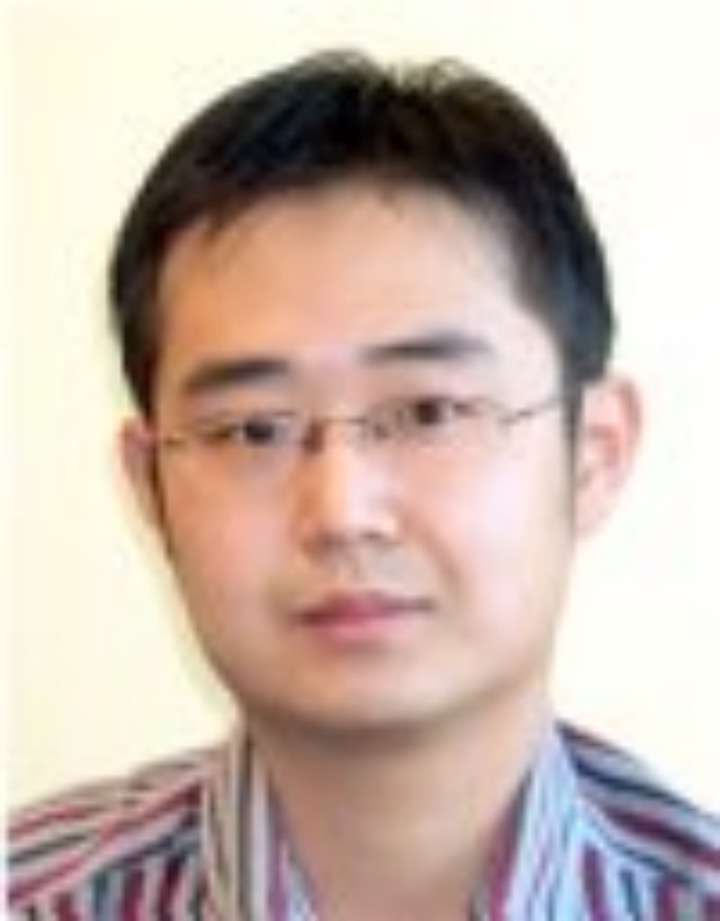}}]{Tom H. Luan}
(Senior Member, IEEE) received the B.E. degree from the Xi'an Jiaotong University, China, the Master degree from the Hong Kong University of Science and Technology, Hong Kong, and the Ph.D. degree from the University of Waterloo, Canada, all in Electrical and Computer Engineering. During 2013 to 2017, Dr. Luan was a Lecturer in Mobile and Apps at the Deakin University, Australia. Since 2017, he is with the School of Cyber Engineering in Xidian University, China, as a professor. Dr. Luan’s research mainly focuses on the content distribution and media streaming in vehicular ad hoc networks and peer-to-peer networking, and protocol and security design and performance evaluation of wireless cloud computing, edge computing and digital twin. Dr. Luan has published over 170 peer reviewed papers in journal and conferences, including IEEE TON, TMC, TMM, TVT and Infocom. He won the 2017 IEEE VTS Best Land Transportation Best Paper award and IEEE ICCS 2018 best paper award. 
\end{IEEEbiography}

\begin{IEEEbiography}[{\includegraphics[width=1in,height=1.25in,clip,keepaspectratio]{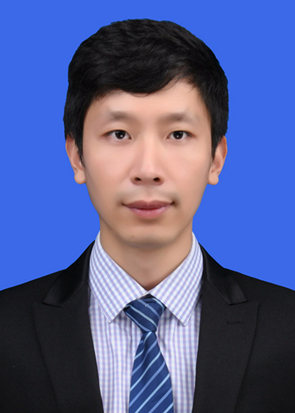}}]{Chengzhe Lai}
 \setlength{\parskip}{0pt}
(Member, IEEE) received his B.S. degree in information security from Xi'an University of Posts and Telecommunications in 2008 and a Ph.D. degree from Xidian University in 2014. He was a visiting Ph.D. student with the Broadband Communications Research (BBCR) Group, University of Waterloo from 2012 to 2014. At present, he is with Xi'an University of Posts and Telecommunications and National Engineering Research Center for Secured Wireless, Xi'an, China. His research interests include wireless network security and privacy preservation.
\end{IEEEbiography}

\begin{IEEEbiography}[{\includegraphics[width=1in,height=1.25in,clip,keepaspectratio]{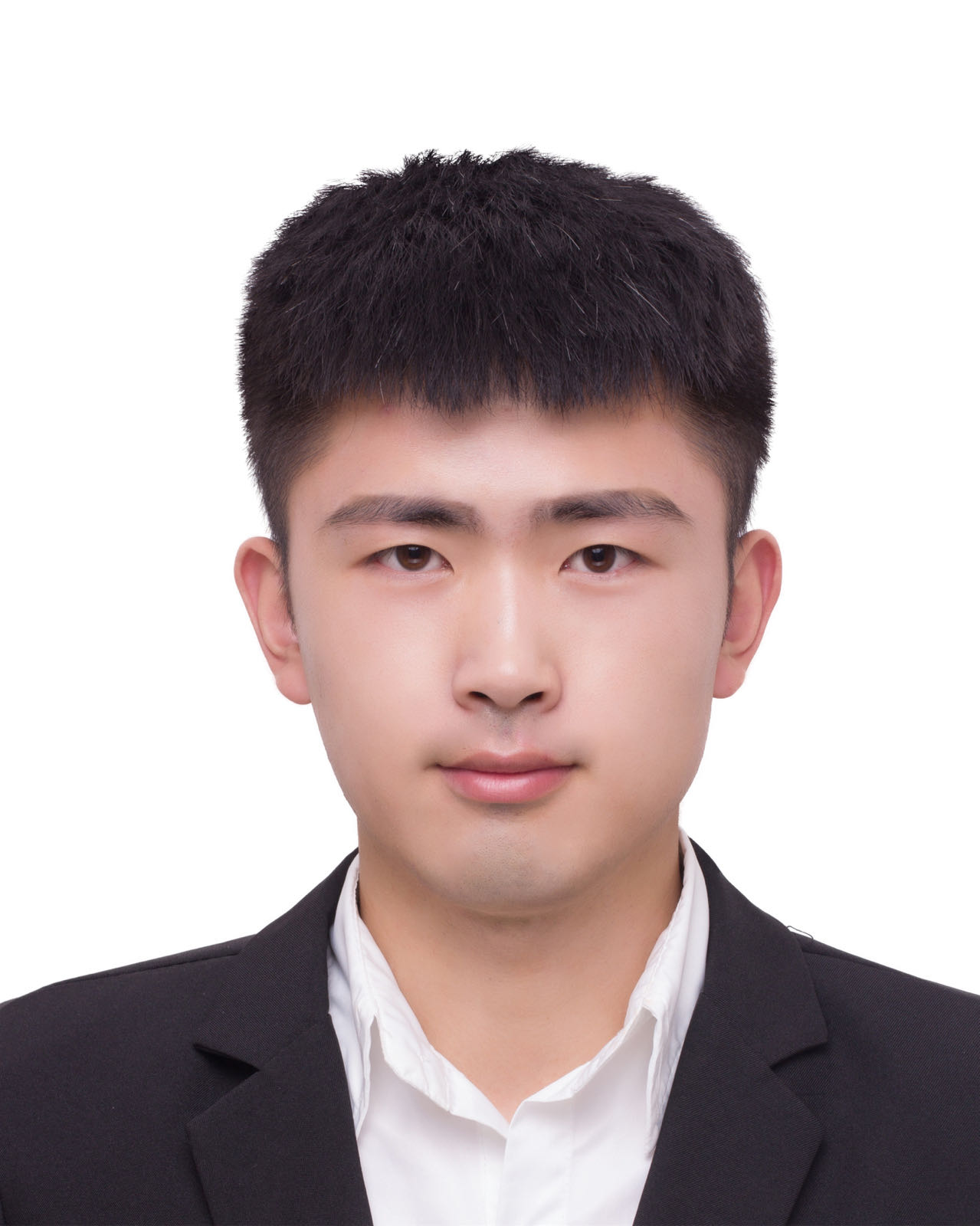}}]{Jinkai Zheng}
 \setlength{\parskip}{0pt}
(Member, IEEE) received his Ph.D. degree in Cyberspace Security from Xidian University, China, in 2024. From 2023 to 2024, he visited PanLab, the Department of Computer Science, University of Victoria, Victoria, BC, Canada. He is currently a Post-Doctoral Researcher with the School of Electrical Engineering and Intelligentization, Dongguan University of Technology, China, and also with the School of Cyber Science and Engineering, Xi’an Jiaotong University, China. His research interests currently include the Internet of Vehicles, Digital Twins, reinforcement learning, and game theory.
\end{IEEEbiography}

\begin{IEEEbiography}[{\includegraphics[width=1in,height=1.25in,clip,keepaspectratio]{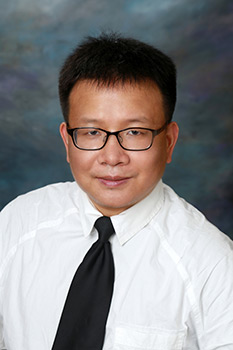}}]{Rongxing Lu}
 \setlength{\parskip}{0pt}
(Fellow, IEEE) received the PhD degree from the Department of Electrical and Computer Engineering, University of Waterloo, Waterloo, ON, Canada, in 2012. He was an assistant professor with the School of Electrical and Electronic Engineering, Nanyang Technological University, Singapore, from 2013 to 2016. He has been an associate professor with the Faculty of Computer Science (FCS), University of New Brunswick (UNB), Fredericton, NB, Canada, since 2016. He was a postdoctoral fellow with the University of Waterloo, from 2012 to 2013. He was a recipient of the Governor Generals Gold Medal for his PhD degree from the Department of Electrical and Computer Engineering, University of Waterloo, the 8th IEEE Communications Society (ComSoc) AsiaPacific Outstanding Young Researcher Award, in 2013, and the 2016 to 2017 Excellence in Teaching Award from FCS, UNB. He currently serves as the vicechair (Publication) of IEEE ComSoc CIS-TC.
\end{IEEEbiography}

\end{document}